\documentclass[fleqn,usenatbib]{mnras}

\usepackage{newtxtext,newtxmath}
\usepackage[T1]{fontenc}
\usepackage{ae,aecompl}

\usepackage{graphicx}	% Including figure files
\usepackage{amsmath}	% Advanced maths commands
\usepackage{enumerate}

\newcommand{\dfourk}{\ensuremath{\mathrm{D4000_n}}}
\newcommand{\hda}{\ensuremath{\mathrm{H\delta_A}}}
\newcommand{\mum}{\ensuremath{\mathrm{\mu m}}}
\newcommand{\ssfr}{\ensuremath{\mathrm{sSFR}}}
\newcommand{\msun}{\ensuremath{\mathcal{M}_\mathrm{\odot}}}
\newcommand{\mhalo}{\ensuremath{\mathcal{M}_\mathrm{h}}}
\newcommand{\rbcg}{\ensuremath{R_\mathrm{bcg}/R_{200}}}

\newcommand{\nmem}{\ensuremath{N_\mathrm{member}}}
\newcommand{\mstar}{\ensuremath{\mathcal{M}_\star}}

\newcommand{\halpha}{H{$\alpha$}}

\title[Inside-out quenching in clusters]{Cluster environment quenches the star formation of low-mass satellite galaxies from the inside-out}

\author[B. Wang et al.]{
Bitao Wang\thanks{E-mail: bt-wang@pku.edu.cn}
\\
Kavli Institute for Astronomy and Astrophysics, Peking University, Beijing 100871, China
}

\date{Accepted 2022 August 25. Received 2022 August 25; in original form 2022 May 9}

\pubyear{2022}

\begin{document}
\label{firstpage}
\pagerange{\pageref{firstpage}--\pageref{lastpage}}
\maketitle

\begin{abstract}
  Environment plays a critical role in the star formation history of galaxies.
  Tidal and hydrodynamical stripping, prominent in cluster environment, can remove the peripheral gas of galaxies and star formation may thus be environmentally suppressed from the outside-in.
  We revisit the environmental dependence of the radial gradient of specific star formation rate (sSFR) profile.
  We probe the radial gradient by using the archival spectral indices \dfourk\ and \hda\ measured from SDSS fiber spectra, to indicate central sSFR, and the total sSFR from fitting the spectral energy distribution.
  Despite the low spatial resolution, the wealth of SDSS data allows to disentangle the dependences on stellar mass, sSFR, and environment.
  We find that low-mass satellite galaxies in the mass range $9<\mathrm{log}\,\mathcal{M}_{\star}/\mathcal{M}_{\odot}<9.8$ on average quench in more inside-out pattern compared to isolated galaxies matched in mass, sSFR, and fiber coverage.
  This environmental effect is particularly strong for galaxies below the star formation main sequence, and peaks for those in the core of massive clusters where the phase-space diagram reveals clear links between the inside-out quenching and orbital properties.
  Our results suggest that both tidal and hydrodynamical interactions in cluster environment suppress the star formation of satellites mainly from the inside-out.
  As accreted gas of low angular momentum from hot gas halos is an important source for replenishing central gas reservoir, we discuss how gas stripping in clusters may lead to starvation and cause inside-out quenching when the outer star-forming discs are not significantly affected.
\end{abstract}

\begin{keywords}
Galaxy: general - Galaxy: formation - galaxies: groups: general - galaxies: star formation
\end{keywords}

%%%%%%%%%%%%%%%%%%%%%%%%%%%%%%%%%%%%%%%%%%%%%%%%%%

%%%%%%%%%%%%%%%%% BODY OF PAPER %%%%%%%%%%%%%%%%%%

\section{Introduction}
\label{sec:intro}
In the local universe, the density of galaxies spans several orders of magnitude, from $\sim0.2\,\rho_{0}$ (where $\rho_{0} \sim 10^{-29.7} g/\mathrm{cm}^{3}$ is the mean field density) in sparse void regions all the way up to $\sim100\,\rho_{0}$ in the cores of massive clusters and $\sim1000\,\rho_{0}$ in most compact groups \citep{1989Sci...246..897G}.
A large variety of galaxy properties are observed to correlate with galaxy environments such as star formation or quenched galaxy fraction \citep{2006MNRAS.373..469B, 2008MNRAS.385.1903L, 2010ApJ...721..193P, 2013MNRAS.430.1447K, 2013MNRAS.428.3306W, 2017ApJ...838...87C}, morphology \citep{1978ApJ...226..559B, 1980ApJ...236..351D, 1999ApJ...518..576P, 2009MNRAS.393.1324B, 2011MNRAS.416.1680C}, kinematics \citep{2011MNRAS.416.1680C, 2020MNRAS.495.1958W}, interstellar medium \citep{2011MNRAS.415.1797C, 2012MNRAS.425..273P, 2015MNRAS.453.2399W, 2017MNRAS.466.1275B, 2019MNRAS.483.5409D} and nuclear activity \citep{2004MNRAS.353..713K, 2011MNRAS.418.2043E, 2013MNRAS.430..638S, 2015MNRAS.448L..72S}.
Generally, red galaxies with early-type morphology and little cold gas content tend to populate the inner part of group \footnote{hereafter group refers to the structure where galaxies are bound within one large dark matter halo while it does not indicate the group mass or richness.
Cluster refers to a massive group.} environment while blue, late-type and gas-rich galaxies are mainly found away from crowded regions.

All these apparent links encourage the idea that environment-related processes are an important driver of the galaxy evolution.
Indeed there are abundant pieces of evidence from both observational and theoretical point of view showing the existence of multiple environmental effects (see the review by \citealt{2006PASP..118..517B}).
Sources of these effects can be broadly classified into two types.
The first type is through gravitational interactions with both galaxies and the entire group potential well.
Gravitational tides from neighbours may supply angular momentum to galaxies \citep{1969ApJ...155..393P,1984ApJ...286...38W} and can condition their overall shape \citep{1979MNRAS.188..273B}.
Depending on velocity dispersion within the group, galaxy-galaxy interactions can either have long duration in small groups, such as during preprocessing \citep{2004ogci.conf..341D,2004PASJ...56...29F}, or have higher frequency but short duration in massive clusters, the so-called galaxy harassment \citep{1996Natur.379..613M,1998ApJ...495..139M}.
When the group mass is large, the tidal force exerted by the entire group potential well becomes effective for perturbing group galaxies \citep{1984ApJ...276...26M,1996ApJ...459...82H}.
The second type is through various kinds of hydrodynamic interactions occurring between gaseous components of galaxies and the hot intergalactic medium (hereafter IGM).
Its importance has been suggested ever since when it became clear that hot IGM is ubiquitous among clusters \citep{1977ApJ...215..401M,1977egsp.conf..369O}.
Such type of interaction can happen in various forms, including ram-pressure stripping \citep{1972ApJ...176....1G,2017ApJ...844...48P}, viscous stripping \citep{1982MNRAS.198.1007N,2015ApJ...806..104R} and thermal evaporation \citep{1977Natur.266..501C,2007MNRAS.382.1481N} all of which are able to remove cold gas of galaxies, particularly for the low-mass ones \citep[e.g.,][]{2013AJ....146..124H, 2020MNRAS.494.2090J}.
Several prototypical galaxies under gas stripping in the Virgo cluster are highlighted in a series of works based on radio interferometry \citep{2004AJ....127.3361K, 2007ApJ...659L.115C, 2009AJ....138.1741C, 2012A&A...537A.143V}.
Though originating from different processes, in some cases several mechanisms can have similar effects to galaxies.
One example is galaxy starvation \citep{1980ApJ...237..692L}, in which the loosely bound outer gaseous halos of galaxies are removed by both tidal interactions and ram-pressure stripping preventing further gas accretion \citep{2002ApJ...577..651B}.

It is difficult to discern the relative importance of all these mechanisms in certain environments.
But one consensus reached by the majority of previous studies is that they are more effective on satellite galaxies, i.e. the less massive galaxies that are gravitationally bound by more massive galaxies.
The high-speed relative motion in hot IGM and their shallow potential well both make them more vulnerable to these effects.
Early studies of M31/M32 system \citep[e.g.,][]{1962AJ.....67..471K,1973ApJ...179..423F} and Milky Way/Magellanic clouds system \citep[e.g.,][]{1976ApJ...203...72T,1982MNRAS.198..707L} have been classic paradigm showing such vulnerability of satellites.
The most massive galaxy in the gravitationally bounded system is often called a "central" galaxy.
Analyses of environmental effects are thus commonly undertaken with the satellite and central galaxy dichotomy   \citep[e.g.,][]{2009MNRAS.394.1213W,2012ApJ...757....4P,2013MNRAS.428.3306W}, which is also adopted in this work.

Despite the fact that these environment-related mechanisms are able to partly explain the various correlations with galaxy environment, it is still under debate to what extent they have played a role.
Is there strong causality between environment and various galaxy properties just like what is shown by those superficial correlations?
Or is this apparent link with environment merely a by-product of other more fundamental processes?
This question lies at the heart of the "nature or nurture" problem.
One embodiment of this problem is the controversy over morphology-density relation \citep{1980ApJ...236..351D, 2003MNRAS.346..601G} which was originally thought to be caused by environmental effects.
Following studies argued for the existence of other more important drivers \citep[e.g.,][]{2009MNRAS.393.1324B,2016ApJ...818..180C,2017ApJ...851L..33G,2019MNRAS.485..666B} such as stellar mass, colour and sSFR.
Without doubt we are still not clear how important these environmental effects are.

Useful information comes from studying the environmental dependence of specific star formation rate (sSFR) radial gradient ($\nabla\,\mathrm{sSFR}$), because various mechanisms at work in group environments can affect different parts of the galactic star-forming discs.
For example, ram-pressure stripping is thought to be more efficient at removing loose peripheral atomic hydrogen gas (HI) than affecting inner dense molecular gas disks \citep{2017MNRAS.467.4282M, 2022arXiv220505698Z}, thus probably tending to suppress outer star formation.
While tidal force by cluster potential well can induce gas inflows and boost star formation in galactic central regions \citep[e.g.,][]{1990ApJ...350...89B}.
So, studying environmental dependence of $\nabla\,\mathrm{sSFR}$ helps to figure out what processes in group environment are important in terms of affecting galactic star formation histories.
Or if we eventually find only weak dependence on environment, the effectiveness of those proposed mechanisms should be doubted.

Previous studies along this thread have been carried out using narrow-band $\mathrm{H}\alpha$ imaging \citep[e.g.,][]{2004ApJ...613..851K,2004ApJ...613..866K,2013A&A...553A..91F}, resolved photometry \citep[e.g.,][]{2007ApJ...658.1006M,2008ApJ...677..970W} and more recently integral field spectroscopy \citep[IFS; e.g.,][]{2013MNRAS.435.2903B,2017MNRAS.464..121S,2018MNRAS.476..580S,2019A&A...621A..98C,2019ApJ...872...50L}.
However, these studies have acquired very different and sometimes discrepant knowledge about how star formation distributions of galaxies are affected in group environment.
The conclusions include 1) outside-in truncation of star formation \citep[e.g.,][]{2004ApJ...613..851K,2013A&A...553A..91F,2017MNRAS.464..121S,2019A&A...621A..98C}, 2) preferential suppression of star formation in inner regions \citep[e.g.,][]{2008ApJ...677..970W,2019A&A...621A..98C} and 3) weak or no effect \citep[e.g.,][]{2007ApJ...658.1006M,2013MNRAS.435.2903B,2018MNRAS.476..580S}.
Even when the general conclusions are similar, the signals they found can still be in tension.
For instance, both using IFS data, \citealt{2017MNRAS.464..121S} found outside-in truncation for massive galaxies with stellar mass in the range $10<\mathrm{log}\,\mathcal{M}_{\star}/\mathcal{M}_{\odot}<11$ while the outside-in signal in \citealt{2019A&A...621A..98C} is for less-massive galaxies only ($9<\mathrm{log}\,\mathcal{M}_{\star}/\mathcal{M}_{\odot}<10$), and they found preferential central suppression for massive galaxies.

In this work, we revisit the environmental dependence of the spatial distribution of star formation by combining SDSS fiber spectral indices (for galaxy central region) and global sSFR measurements to indicate the (relative) shape of sSFR\footnote{We approach the profiles of sSFR instead of SFR because characterizing the stellar population by the fraction of newborn stars is more representative of star formation status of galaxies.} profiles.
This brings sufficient statistics to the investigation, which is crucial, because unambiguous environmental dependence can only be extracted when other important factors, such as stellar mass and total star formation level, are properly controlled.
Current IFS samples can still lack such statistics, especially for low-mass galaxies among which the environmental effects are usually the strongest.
Even with currently the largest IFS survey MaNGA \citep{2015ApJ...798....7B}, the sample size is at least an order of magnitude smaller than the sample studied in this work, and would limit the parameter control when we aim to explore in more detail how the sSFR profiles correlate with galaxy environment (see section \ref{subsec:env}).

Throughout this paper we adopt cosmological parameters from WMAP-9 \citep{2013ApJS..208...20B} in which $\mathrm{H}_0=69.3\,\mathrm{km}\,\mathrm{s}^{-1}\,\mathrm{Mpc}^{-1}$, $\Omega_\mathrm{m}=0.286$ and $\Omega_{\Lambda}=0.714$ and a Chabrier IMF.

%%%%%%%%%%%%%%%%%%%%%%%%%%%%%%%%%%%%%%%%%%%%%%%%%%%%%%%%%%%%%%%%%%%%%%%%
\section{Sample}
\label{sec:data}

\subsection{MPA-JHU and GSWLC catalogues}
\label{subsec:cat}

Our galaxy sample is assembled out of the MPA-JHU catalogue and the version 2 of GALEX-SDSS-WISE Legacy Catalogue \citep[GSWLC-2,][]{2016ApJS..227....2S,2018ApJ...859...11S}.

The MPA-JHU catalogue is based on the Sloan Digital Sky Survey Data Release 7 \citep[SDSS DR7,][]{2000AJ....120.1579Y,2009ApJS..182..543A}, providing both spectral and photometric measurements from SDSS as well as value-added derived quantities for more than 800,000 unique galaxies. We heavily use the spectral indices (more details in section \ref{subsec:less}) measured from SDSS spectra which were extracted from fibers of 3 arcsec diameter centered on galaxies. We also take the radius enclosing 50\% of the total r-band Petrosian flux $\mathrm{R_{50}}$ as the apparent angular size of galaxies.

Despite the fact that MPA-JHU catalogue does provide SFR, we use the values from GSWLC-2 instead. GSWLC-2 is a value-added catalogue for SDSS galaxies within the GALEX \citep[Galaxy Evolution Explorer,][]{2005ApJ...619L...1M} footprint.
It provides better SFR measurements in overall by adopting the ultra-violet (UV) data in the multi-band spectral energy distribution (SED) fitting. The UV data is from GALEX, which is a space telescope mapping the sky in two UV bands, FUV (1350-1750 {\rm \AA}) and NUV (1750-2800 {\rm \AA}). Compared with optical SDSS bands, these UV bands are more sensitive to short-lived massive stars, thus to recent star formation.
GSWLC-2 also uses the 22 \mum\ mid-infrared (MIR) band taken by WISE \citep[Wide-field Infrared Survey Explorer,][]{2010AJ....140.1868W}, which is another space telescope providing all sky images in MIR bands. The 22 \mum\ band can trace the absorbed UV light re-emitted by the dust, improving the estimation of recent SFR.
For consistency, we also use the stellar mass from GSWLC-2 which is derived by the same SED fitting procedure.

We use the medium UV depth version of the GSWLC catalogue, taking a balance between the depth of GALEX images and the sky coverage. Our sample thus have a sSFR detection limit of $\mathrm{sSFR} > 10^{-11.7}\,\mathrm{yr^{-1}}$, satisfying the main goal of studying galaxies at low star formation level. The matching between MPA-JHU and GSWLC-M2 is done with a 3 arcsec searching radius, giving a sample of 343,791 galaxies. Changing the matching radius has negligible effect to our sample (differing by less than 0.03\% when matching radius ranges from 1 arcsec to 5 arcsec).

We further constrain our sample with the following criteria:
\begin{equation}
\label{equ:cut}
    \begin{aligned}
        \qquad \qquad \qquad \qquad 0.01&<z<0.085; \\
        14.5&<\mathrm{m}_\mathrm{r}<17.77; \\
        9&<\mathrm{log}\,\mathcal{M}_{\star}<11.5; \\
        0.259&<\mathrm{b/a} \, ,
    \end{aligned}
\end{equation}
where $z$ is redshift, $\mathrm{m}_\mathrm{r}$ is apparent Petrosian magnitude \citep{1976ApJ...209L...1P} in SDSS r-band, $\mathcal{M}_{\star}$ is stellar mass, and $\mathrm{b/a}$ is the ratio between minor and major axis of the 25 $\mathrm{mag}/\mathrm{arcsec}^2$ isophote in SDSS r-band.

The axial ratio cut (equivalent to inclination angle smaller than $75^{\circ}$ for a razor-thin disc) removes edge-on galaxies to avoid large uncertainty in correcting for strong dust extinction.
We limit the redshift below 0.085 as a compromise between sample size and completeness (as also in e.g., \citealt{2006MNRAS.373..469B}).
At $z=0.085$, the SDSS spectroscopic survey is complete for galaxies with absolute r-band magnitude $\mathrm{M}_\mathrm{r}<-19.5$ or stellar mass about $\mathcal{M}_{\star}>10^{10}\,\mathcal{M}_{\odot}$ \citep{2006ApJS..167....1B}.
This magnitude limit is the same as the one adopted for group galaxies defined as halo proxy in the group catalogue used in this work (see section \ref{subsec:yang}), making the halo mass more reliable below $z=0.085$.
Even though the sample is not complete for galaxies with $\mathcal{M}_{\star}<10^{10}\,\mathcal{M}_{\odot}$ out to $z=0.085$, the analyses throughout this work make proper control of stellar mass and sSFR so that the low-mass galaxies in different environment are compared in the same subvolume where they are complete.
The lower redshift limit and the brighter apparent r-band Petrosian magnitude limit are applied to exclude nearby galaxies with too large angular size as their photometry are not properly handled by the SDSS pipeline \citep{2011AJ....142...31B}.
After this cut, our sample size reduces to 119,820.

Our analysis is applied only to galaxies with $\ssfr > 10^{-11.7}\,\mathrm{yr^{-1}}$, the nominal detection limit of the GSWLC-M2 catalogue.
Below this limit, the error in the total SFR surges to 0.7 dex and probing the sSFR radial profile by central spectral indices and total sSFR thus becomes highly uncertain.

\subsection{Galaxy environment}
\label{subsec:yang}

We use the group catalogue constructed by \citet{2012ApJ...752...41Y} to classify the environment of each galaxy. It was built by applying an iterative group finder algorithm to SDSS galaxies. In each iteration the halo properties of the tentative galaxy groups (identified via friends-of-friends algorithm) are computed and then used to update the group membership for next iteration \citep{2007ApJ...671..153Y}. The catalogue associates each galaxy to one galaxy group, hence one dark matter halo as well. Based on this, we classify the galaxies into three categories: central, satellite and isolated galaxies. Centrals and satellites are the members of multi-member groups, with the former to be the most massive one. The isolated galaxies belong to the groups with only one member.

The catalogue also provides dark matter halo mass estimation, based on the total stellar mass or luminosity of bright group members (absolute r-band magnitude $\mathrm{M}_\mathrm{r}<-19.5$) via abundance matching. A mock test suggests its typical uncertainty is about 0.3 dex \citep{2012ApJ...752...41Y}. The halo mass links with a certain virial radius of the halo $R_{200}$:

\begin{equation}\label{r200}
\qquad \qquad \qquad \mathrm{R}_{200}=\Bigg[\frac{\mathcal{M}_{200}}{\frac{4\pi}{3}200\Omega _\mathrm{m} \frac{3\mathrm{H}_0^2}{8\pi \mathrm{G}}}\Bigg]^{\frac{1}{3}}\,\,(1+z)^{-1}.
\end{equation}

Among the several catalogues with slightly different redshift completeness, we take the group catalogue constructed with SDSS redshifts only, which contains 599,301 galaxies.
Using the other versions makes negligible difference.
After matching with the group catalogue, we get a sample of 112,028 galaxies.

% %#################################################################

\begin{figure*}
	\begin{center}
    \includegraphics[width=0.48\textwidth]{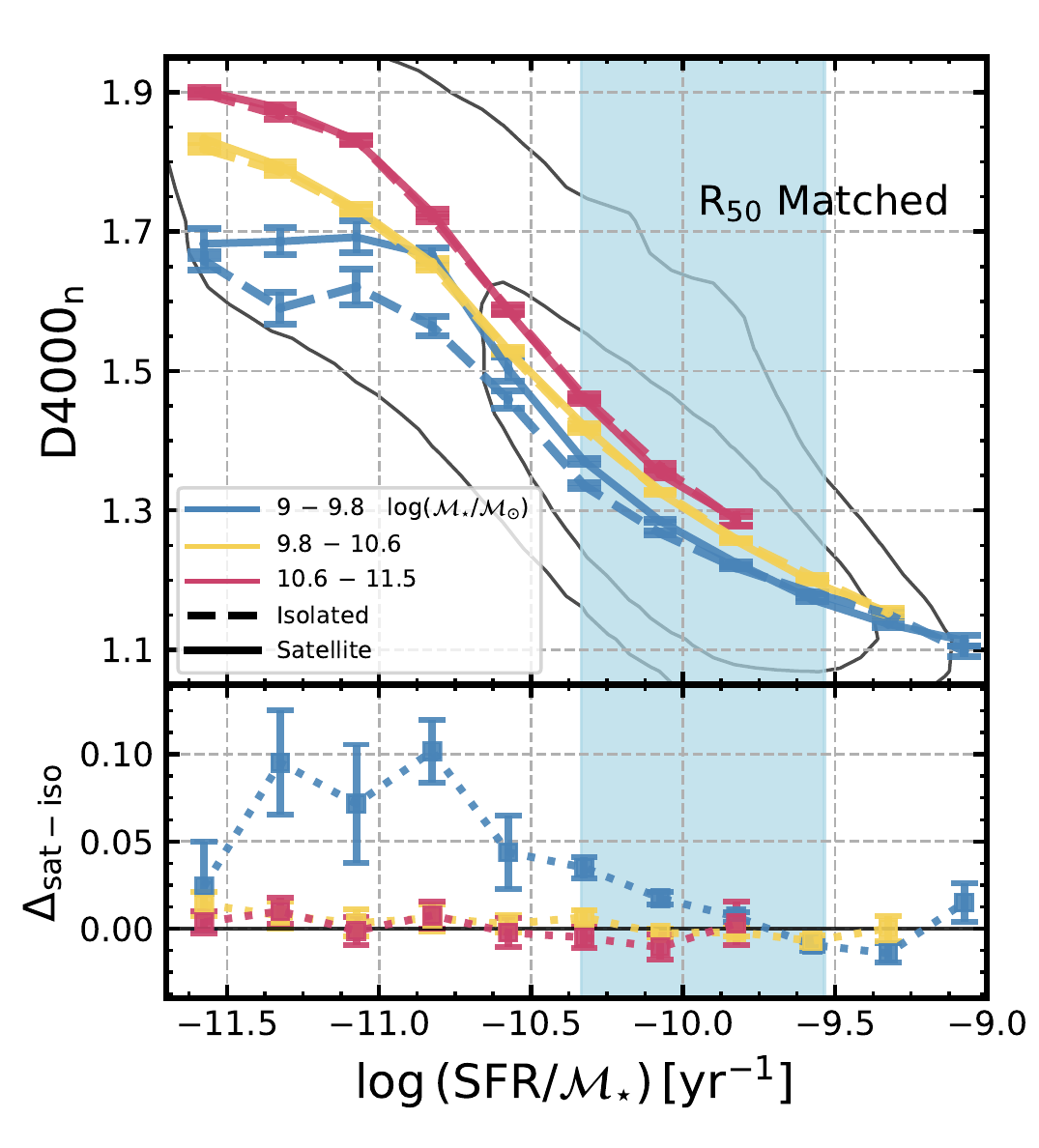}
    \includegraphics[width=0.48\textwidth]{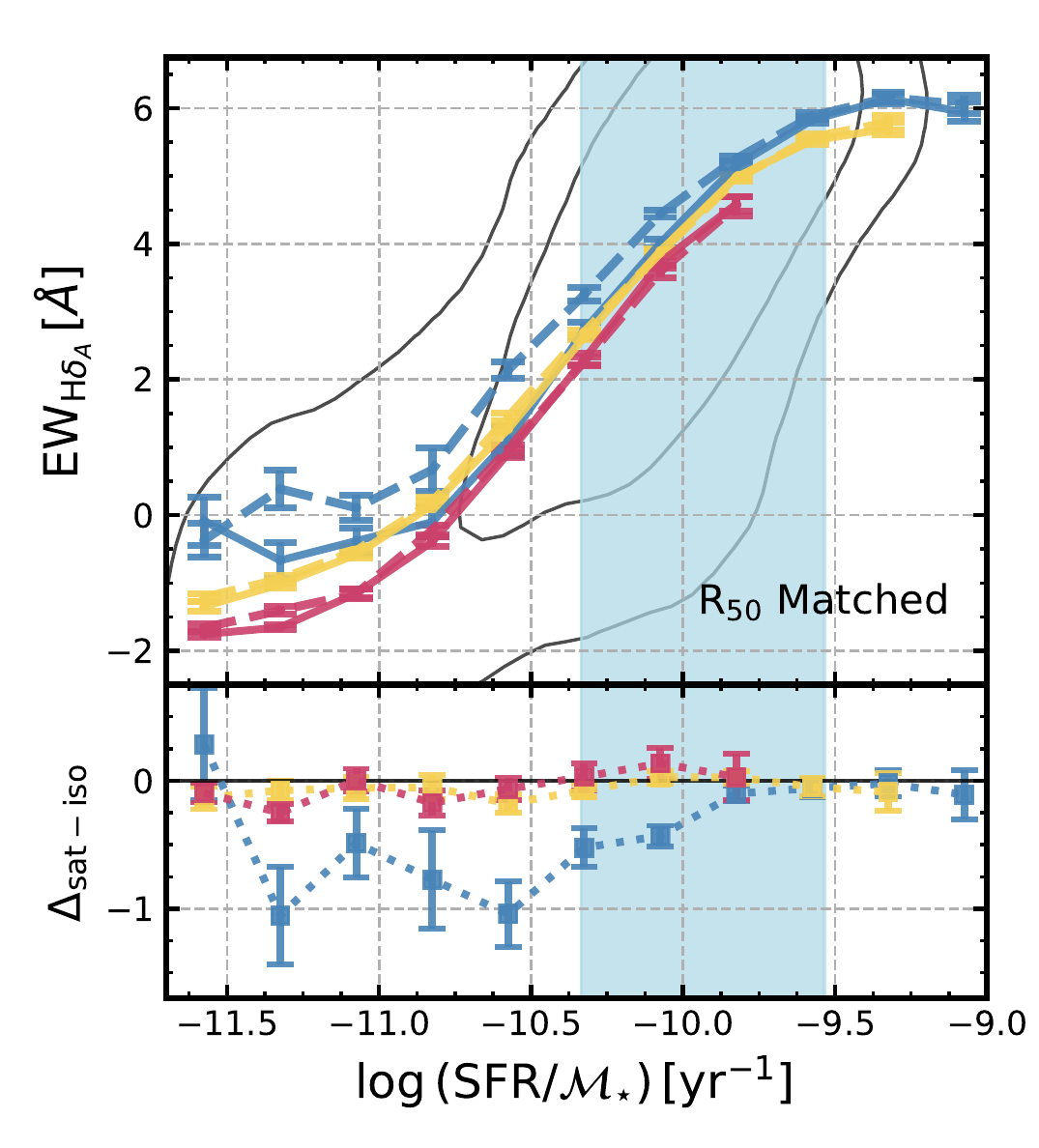}
    \caption{Top panel: \dfourk\ (left) and \hda\ (right) as a function of sSFR for isolated (dashed line) and satellite galaxies (solid line), divided into three stellar mass bins (red, yellow and blue colours). The isolated and satellite galaxies are matched with $\mathrm{R_{50}}$. The background grey contour is derived from kernel density estimation with $V_\mathrm{max}$ correction, enclosing 68\% and 95\% of probability for galaxies in stellar mass range $10^{9}-10^{11.5}\,\msun$ and sSFR range $10^{-11.7}-10^{-9}\,\mathrm{yr}^{-1}$. Blue shaded region marks the span of the SFMS for galaxies in the lowest mass bin. Bottom panel: the difference between satellite and isolated galaxies.
    The bins with less than 20 galaxies are discarded.}
    \label{fig:100bs1}
	\end{center}
\end{figure*}

\section{Results}
\label{sec:result}

\subsection{Suppressed star formation in the center of satellite galaxies}
\label{subsec:less}

The SDSS single-fiber spectra are extracted from the central part of the galaxies, within a physical radius of 0.3, 1.5, 2.4 kpc respectively at $z=0.01,0.05,0.085$, where 0.05 is about the mean redshift of our sample.
We use the \dfourk\ and the Balmer absorption feature \hda\ to indicate the central sSFR (see also \citealt{2004MNRAS.353..713K}).
\dfourk\ is a break feature at around $4000$ {\rm \AA} mainly due to a series of metal absorption lines on the blueward side of $4000$ {\rm \AA}.
These lines are most prominent for stars with spectral types later than K \citep{1985ApJ...297..371H}, i.e. old stellar populations.
While the opacity at Balmer line \hda\ peaks among young massive stars with spectral types around A.
Therefore if galaxies are more dominated by young stars (i.e. high sSFR), \hda\ and \dfourk\ are respectively higher and lower.
These two indices are insensitive to dust extinction as they are flux ratios in adjacent and narrow spectral windows.
This is particularly important because the central regions of galaxies are usually highly dust obscured which may introduce large uncertainty in the measured sSFR \citep{2017MNRAS.469.4063W}.
With the central sSFR indicated by SDSS spectral indices and total sSFR from SED fitting, it becomes possible to roughly probe the gradient of the sSFR radial profiles.
Though the central and total sSFR are not measured in a consistent way, we prove in Appendix \ref{app:fea} the feasibility in a statistical sense with a smaller sample of galaxies with IFS data.

We investigate the environmental dependence of the relative difference in sSFR radial gradient by comparing the central sSFR of satellite and isolated galaxies at fixed total sSFR and stellar mass.
To ensure that the fiber measurements are on similar scales, we match the apparent angular size of galaxy $\mathrm{R_{50}}$ so that fibers cover similar fractions of galaxy total light.
An alternate aperture controlling is to match redshift, to make fibers cover the same physical scales.
We have tested and found that the two ways lead to the same conclusion.

Specifically, in a certain bin of stellar mass and total sSFR, we minimally trim the satellite and isolated galaxy samples to reach the same $\mathrm{R_{50}}$ distribution in 0.2 arcsec resolution (i.e. getting the maximally overlapping distribution).
The trimming is done in every $\mathrm{R_{50}}$ bin by sampling with replacement a same number (i.e. minimum of $[\mathrm{N_{sat},N_{iso}}]$) of the isolated and satellite galaxies.
We repeat this matching process for 1000 times to estimate the statistical uncertainty in distribution moments (see also \citealt{2008MNRAS.385.1903L} and \citealt{2015MNRAS.448L..72S}).
We compute the median \dfourk\ and \hda\ for each matched isolated and satellite sample respectively, and the mean and the standard deviation of the 1000 values are taken as the final measurement and its uncertainty.

In Fig. \ref{fig:100bs1}, we show the relation between the central sSFR, indicated by \dfourk\ (left panel) and \hda (right panel), and the total sSFR for satellite and isolated galaxies matched in $\mathrm{R_{50}}$.
At given total sSFR, more massive galaxies have lower central sSFR (i.e. higher \dfourk\ and lower \hda).
It is consistent with the well established observation that massive galaxies generally show more positive sSFR profiles \citep[e.g.,][]{2016ApJ...819...91P,2018MNRAS.474.2039E,2018ApJ...856..137W}.

Noteworthily, in the lowest mass bin and at given total sSFR, satellite galaxies show prominently higher \dfourk\ (hereafter we term this "the central \dfourk\ excess") when compared to their isolated counterparts (left panel).
This signal of environmental dependence of sSFR radial gradients is strongest when the total sSFR of galaxies is well below the star formation main sequence (SFMS; shown by the blue shaded region), whose location is defined by the peak sSFR at given mass of the volume-corrected number density distribution of our sample galaxies (see also the Appendix A of \citet{2020MNRAS.495.1958W}).
Similar trend is also spotted in \hda\ versus sSFR diagram (right panel), where the low-mass satellite galaxies have systematically lower \hda\ values.
This suggests that environmental effects preferentially suppress the central star formation of galaxies, making the sSFR profile gradient more positive in a relative sense.
Conclusion remains the same if we use total SFR derived from a different recipe, for example measured directly from UV and MIR luminosity (see Appendix \ref{app:nuvmir}).

\subsection{The dependence on galaxy environment}
\label{subsec:env}

In this section we further explore what suppresses the central star formation in low-mass satellite galaxies by studying how the \dfourk\ excess correlates with galaxy environment.
We investigate this environmental dependence in two sSFR windows: $10^{-10.4}-10^{-9.4}\,\mathrm{yr}^{-1}$ and $10^{-11.4}-10^{-10.4}\,\mathrm{yr}^{-1}$.
These two windows respectively cover normal star-forming galaxies around the SFMS, and galaxies below the SFMS but still with detectable star formation activity.
Our galaxies in the low sSFR bin have a median NUV-r colour index of $\sim4$ which falls onto the conventional green valley on the colour-magnitude diagram (e.g., as in \citealt{2007ApJS..173..267S}).
We use three parameters to quantify environment of satellites: the halo mass of the group \mhalo, the normalized projected distance to the central galaxy \rbcg\ (which is effectively the distance to the halo center\footnote{In groups with few members the weighted-geometric center can be a better tracer of the bottom of the group potential well, as there may not be a dominating central galaxy. We have tested for small groups using this alternate definition of group center and found consistent results that leave our conclusion unchanged.}) and the group richness \nmem\ (i.e. number of galaxies within the group).

\begin{figure*}
	\begin{center}
		\includegraphics[width=0.95\textwidth]{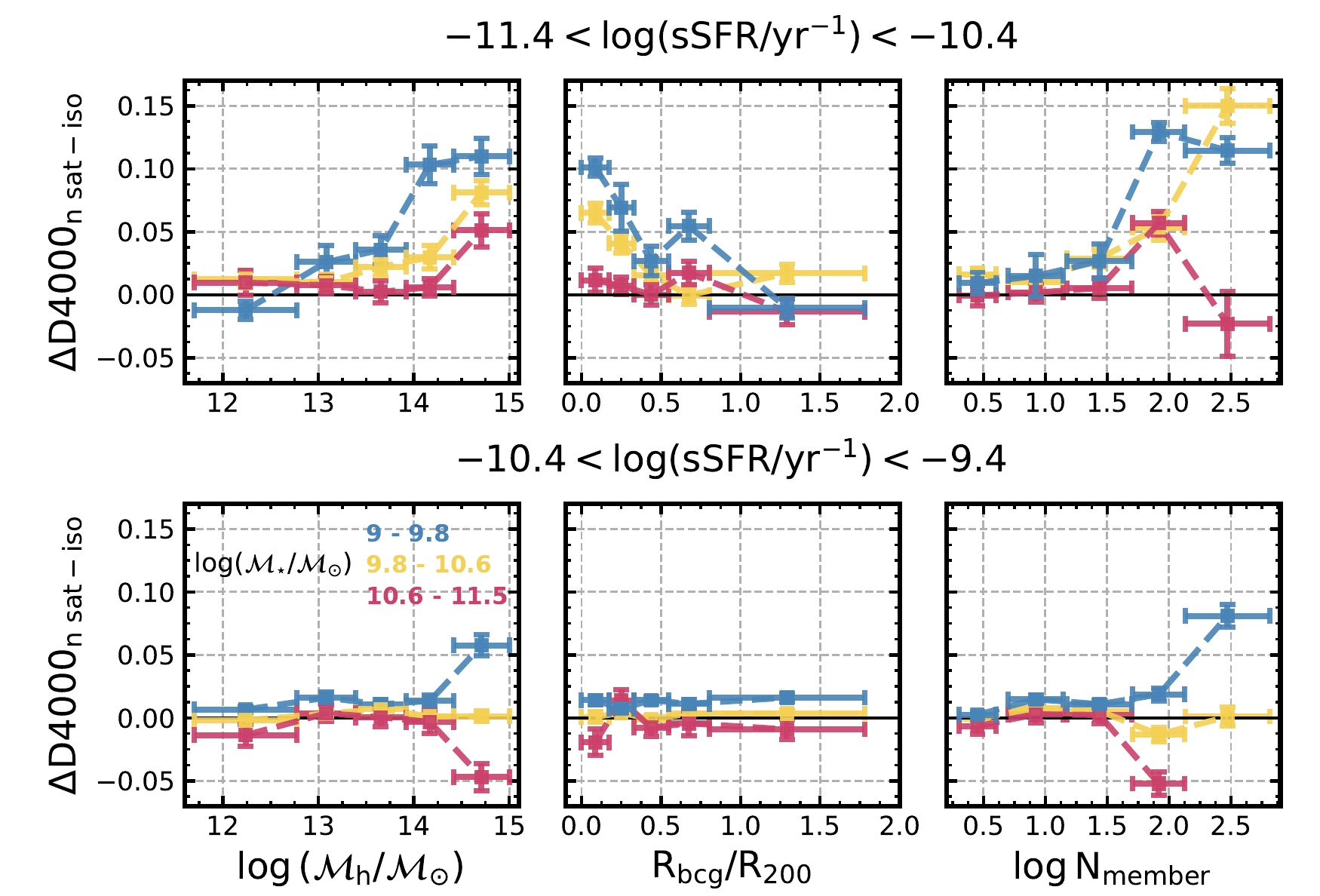}
    \caption{
        Central \dfourk\ excess as a function of the halo mass (\mhalo, left panel), the normalized projected distance to the central galaxy (\rbcg, middle panel) and the group richness (\nmem, right panel). Galaxies are split into three stellar mass bins (blue, yellow and red lines), and two sSFR ranges (top and bottom panels), respectively. The horizontal error bar shows the width of the parameter bin, and the vertical error bar shows the statistical uncertainty estimated with bootstrapping technique. The bins with less than 20 galaxies are discarded.
		}
		\label{fig:Edep}
	\end{center}
\end{figure*}

Fig. \ref{fig:Edep} represents the \dfourk\ excess as function of these group properties in low and high sSFR bins. The \dfourk\ excess is again calculated by comparing satellite galaxies and their matched isolated counterparts with $\Delta\log(\mstar) < 0.1$, $\Delta\log(\mathrm{sSFR}) < 0.1$ and $\Delta \mathrm{R_{50}} < 0.2\,\mathrm{arcsec}$.
For galaxies with low sSFR, the \dfourk\ excess apparently correlates with all environment properties.
The satellite galaxies have redder cores (i.e. more suppressed central star formation) when they are: 1) in more massive halos; 2) closer to the center of galaxy groups; 3) in groups with more members.
The correlation steepens toward lower stellar mass.

For galaxies in the high sSFR bin, the environmental dependence is much weaker.
Clear \dfourk\ excess only exists in the largest \mhalo\ and \nmem\ bins, and only for low-mass galaxies.
We note that for massive galaxies with high sSFR shown by the red lines in the bottom panels, in the most massive groups, the \dfourk\ signal is not excess but deficiency, indicating enhanced central star formation compared with galaxies in the field environment.

\begin{figure*}
	\begin{center}
		\includegraphics[width=0.99\textwidth]{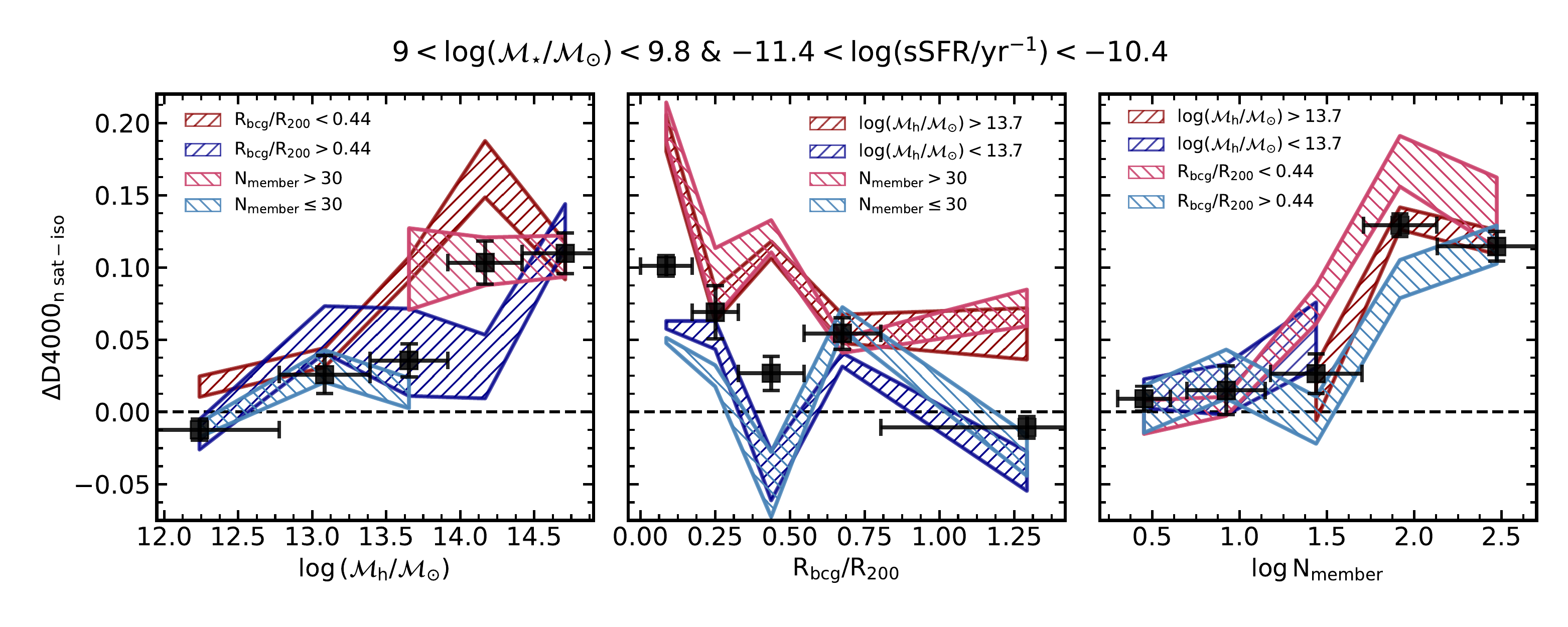}
		\caption{
		The same as in Fig. \ref{fig:Edep} but only for low-mass low-sSFR galaxies (black squares). The sample is further split into sub-samples with different environmental parameters (red and blue stripes, see legend in each panel).
		}
		\label{fig:Edep1}
	\end{center}
\end{figure*}

To further break down the environmental dependences of the \dfourk\ excess of low-mass satellites of low sSFR, in Fig. \ref{fig:Edep1} we apply more environment control to the correlation between the \dfourk\ excess and certain environment properties.
In the first panel, we show the central \dfourk\ excess as a function of \mhalo\ in bins of high/low \rbcg\ and \nmem\ respectively (split by the median value, i.e. 0.44 and 30, of the low-mass and low-sSFR satellite sample).
The second and third panel show the other two dependences on \rbcg\ and \nmem\ with further environment control in a similar manner.
We note that there are 88 individual massive groups included in the $\mathcal{M}_h>10^{13.7}\,\mathcal{M}_{\odot}$ bin, making the result in this bin statistically representative for large groups.
The relations for the low-mass and low-sSFR satellites without further environment control in Fig. \ref{fig:Edep} are shown for reference by black symbols.

We find that \mhalo\ and \nmem\ are almost interchangeable.
In the first panel, the relations of \dfourk\ excess and \mhalo\ in bins of high/low \nmem\ (light red and light blue) are just the general relation (black symbols) at higher and lower \mhalo\ end.
The same case is seen in the third panel, and in the second panel the binning by \mhalo\ or \nmem\ gives the same relations.
This is resulted from the tight correlation between \mhalo\ or \nmem\ and among satellite galaxies with non-zero host halo mass catalogued in \citealt{2012ApJ...752...41Y}, the Spearman rank correlation coefficient between \mhalo\ or \nmem\ is as high as 0.92.

Comparing between the left two panels, generally the \dfourk\ excess depends more on \mhalo\ than on \rbcg\ .
The \dfourk\ excess is small in less massive halos, nearly irrespective of groupcentric radius.
This is shown by the overlapping dark red and dark blue bands at low \mhalo\ end in the first panel and also the relatively flat relation in the second panel (dark blue band).
\dfourk\ excess is present in massive halos even at very large \rbcg\ .
The dependence on \rbcg\ starts to become significant in massive halos, especially at the center where we observe \dfourk\ excess as high as 0.2.
These results together seem to suggest the first-order importance of halo mass and also that the physical mechanism gets strongly enhanced in cluster center.

\begin{figure*}
	\begin{center}
		\includegraphics[width=0.45\textwidth]{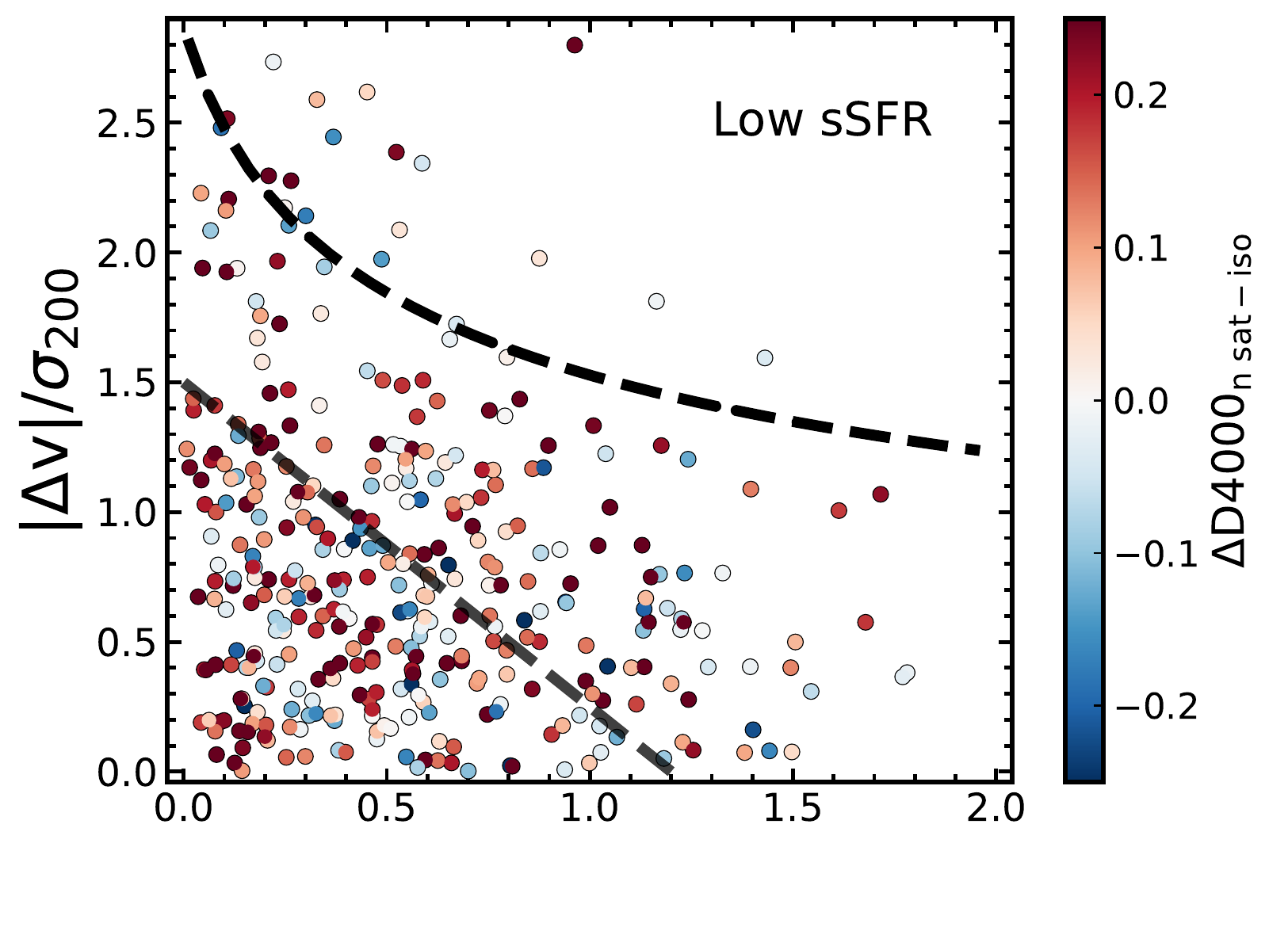}
		\includegraphics[width=0.45\textwidth]{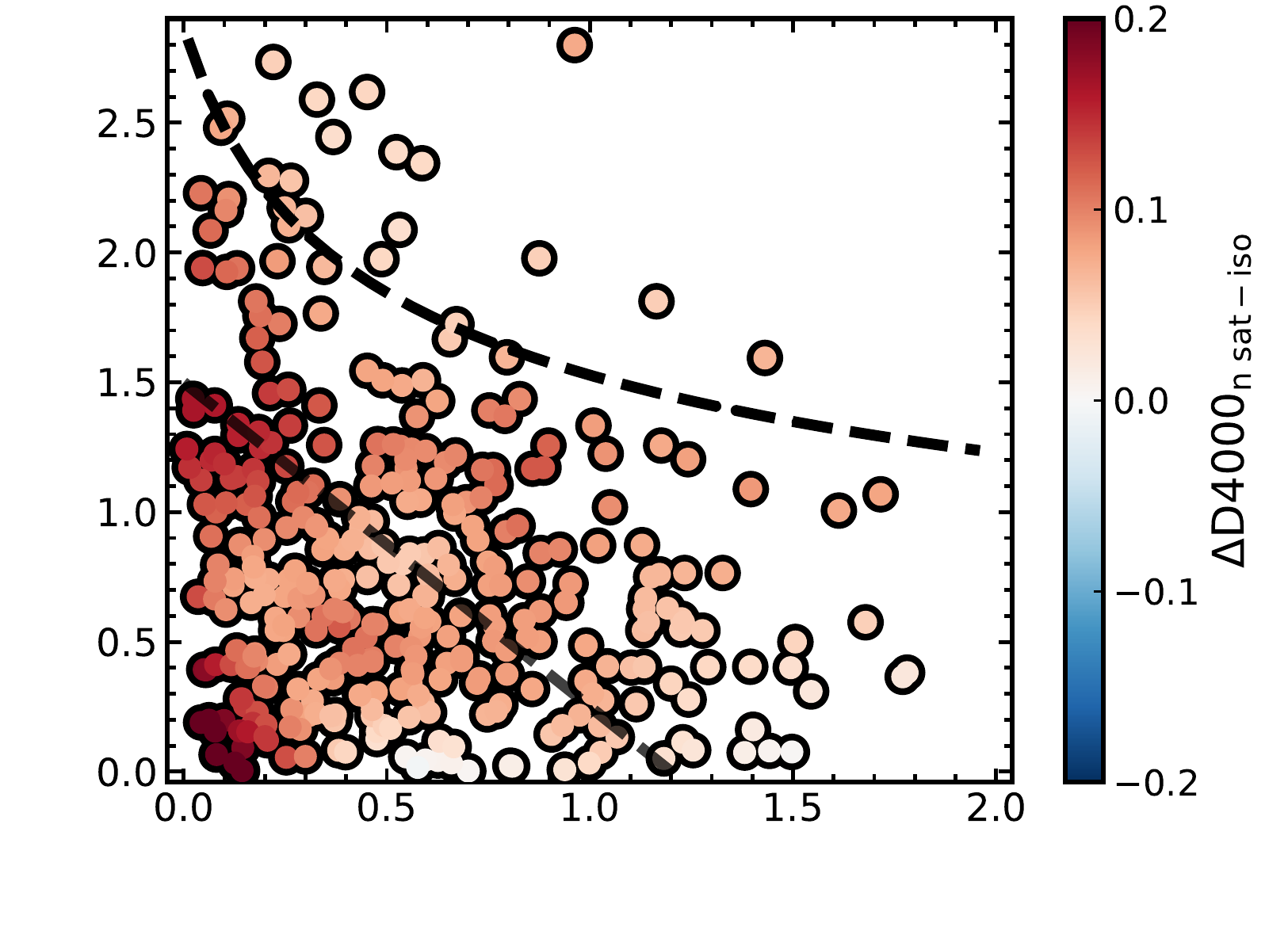} \\
		\includegraphics[width=0.45\textwidth]{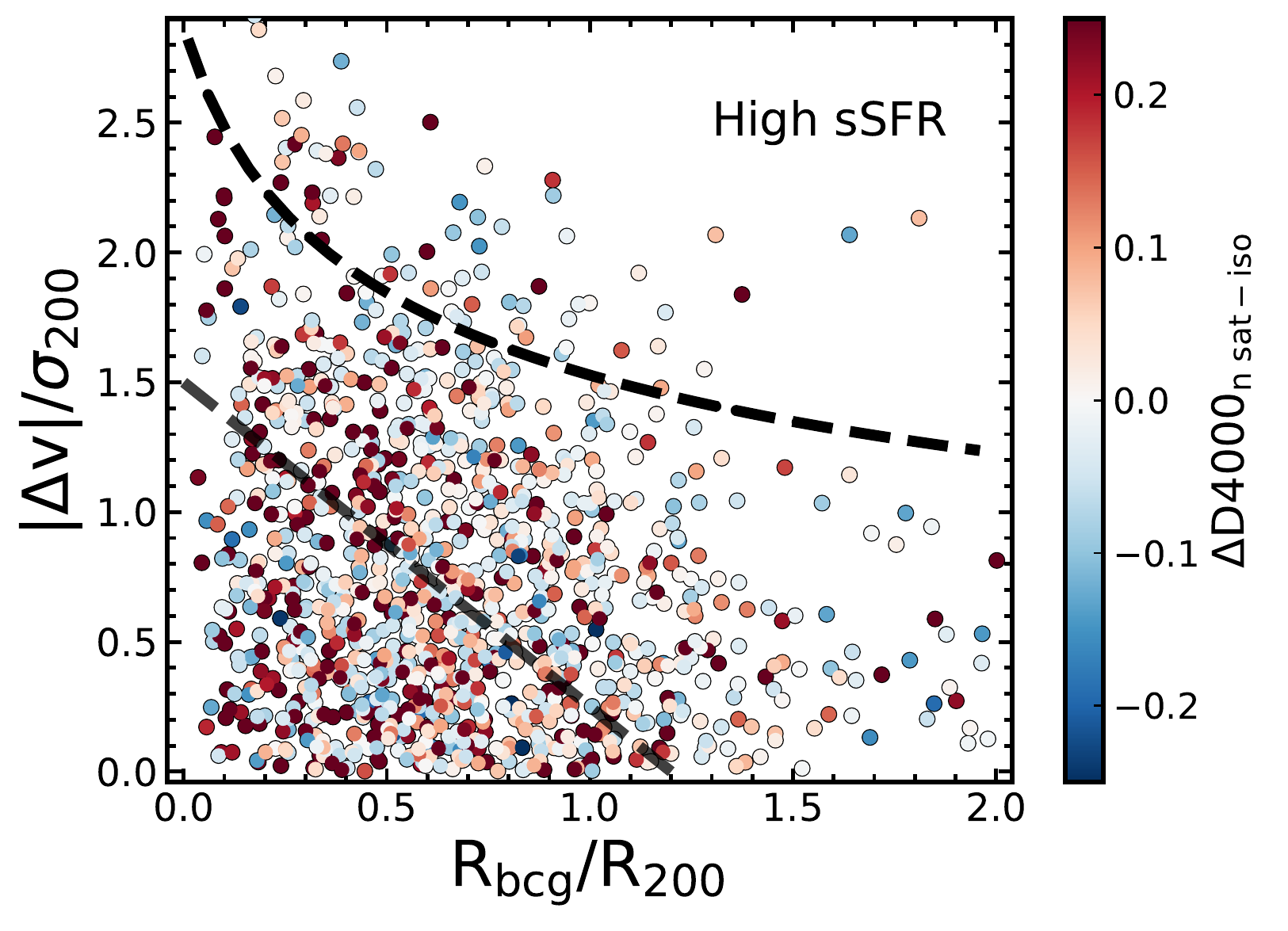}
		\includegraphics[width=0.45\textwidth]{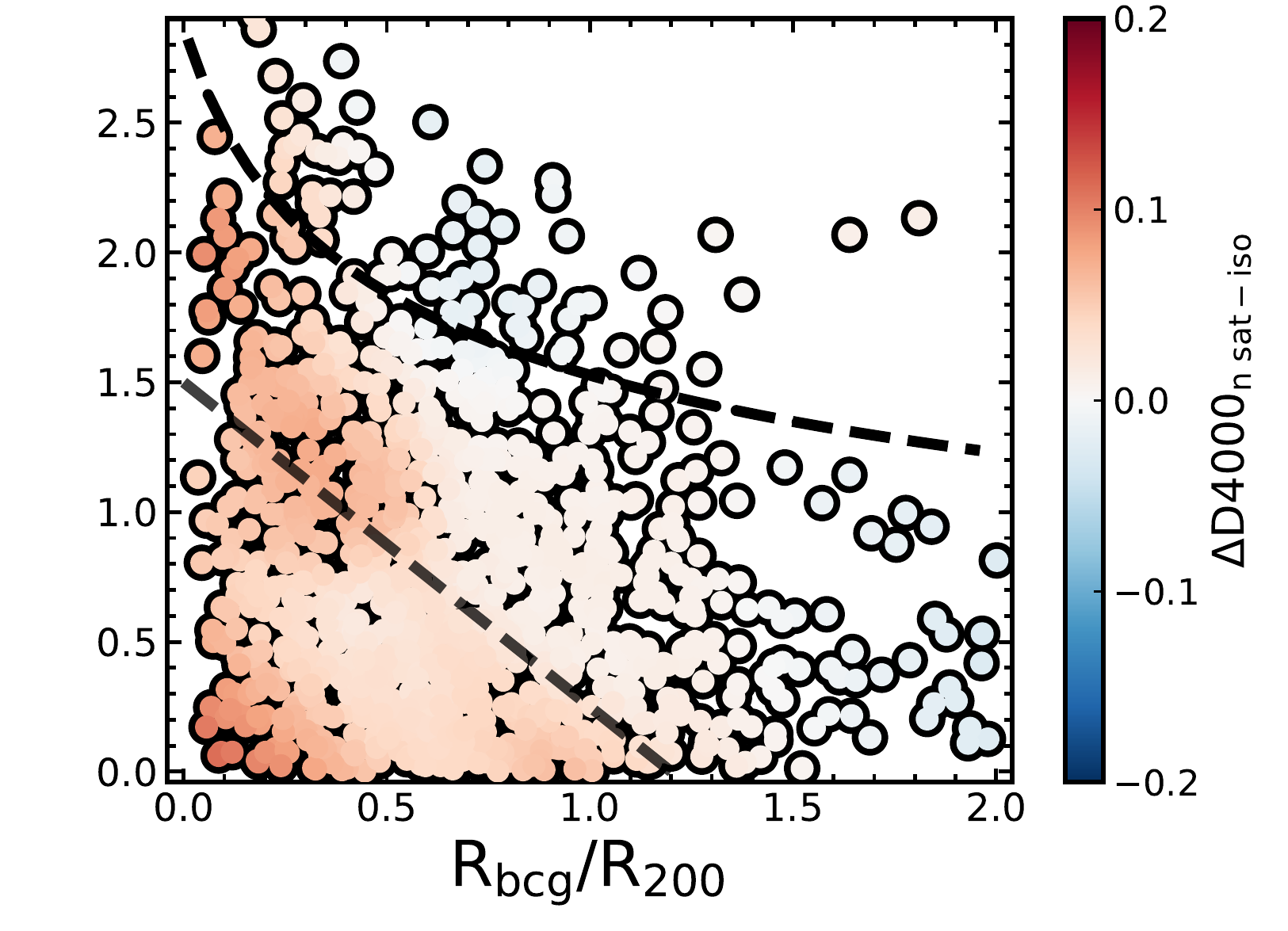}
		\caption{
		    The central \dfourk\ excess on the phase-space diagram for the low-mass galaxies ($10^9 - 10^{9.8}\,\msun$) in massive halos ($\mhalo > 10^{13.7}\,\msun$). y-axis is the line-of-sight velocity of satellites relative to their host clusters, normalized by cluster velocity dispersion.
        The upper and lower row are for galaxies in the low and high sSFR range as in Fig. \ref{fig:Edep}.
        The right panels are locally averaged version of the left panels, showing the underlying trend.
        Satellites in the lower triangle region are considered a virialized part of the clusters.
        The black dashed curve is the projected escape velocity normalized by cluster velocity dispersion and galaxies far beyond the curve are not gravitationally bound by the clusters.
		}
		\label{fig:psd}
	\end{center}
\end{figure*}

Taking a step further we introduce the relative velocity of satellites into the analysis to try to link the central \dfourk\ excess to the dynamic status of satellites in their host halos.
Fig. \ref{fig:psd} shows low-mass satellites ($10^9-10^{9.8}\,\mathcal{M}_{\odot}$) in massive halos ($\mathcal{M}_h>10^{13.7}\,\mathcal{M}_{\odot}$) on the phase-space diagram \citep[i.e. normalized relative velocity versus normalized projected distance; See also][]{2015MNRAS.448.1715J}.
We calculate the absolute difference of line-of-sight velocities between the satellite and cluster as $|\Delta v| = c|z-z_c|/(1+z_c)$ where $z_c$ is the luminosity weighted redshift of cluster member galaxies.
The velocity difference is then normalized by the cluster velocity dispersion $\sigma_{200}$ (equation 6 of \citealt{2007ApJ...671..153Y}).

We mark the boundary of virialized area by a black straight line, below which galaxies are approximately within the part of the cluster in dynamical equilibrium.
The black dashed curve represents the normalized projected escape velocity $v_\mathrm{esc}/\sigma_{200}$ based on a Navarro-Frenk-White halo \citep{1996ApJ...462..563N} of concentration $c_\mathrm{NFW}=6$.
Starting from the mass profile of a halo one can calculate the potential and thus the escape velocity:
\begin{equation}\label{equa:vesc}
\qquad
v_\mathrm{esc,3D}=\sqrt{\frac{2GM_{200}}{R_{200}}\times g(c_\mathrm{NFW}) \times \frac{ln(1+c_\mathrm{NFW}x)}{x}}
\end{equation}
where
\begin{equation}
\qquad
g(c_\mathrm{NFW})=\Big [ ln(1+c_\mathrm{NFW})-\frac{c_\mathrm{NFW}}{1+c_\mathrm{NFW}} \Big ] ^{-1}
\end{equation}
and
\begin{equation}
\qquad
x=r_\mathrm{3D}/R_{200}
\end{equation}
We project velocity along the line of sight and project the distance on the sky plane using the average relations $v_\mathrm{esc} = \frac{1}{\sqrt{3}}v_\mathrm{esc,3D}$ and $r = \frac{\pi}{4}r_\mathrm{3D}$.

The same as previous analyses we match the satellites by isolated galaxies of stellar mass, sSFR and $R_{50}$ differences less than 0.1 dex, 0.1 dex and 0.2 arcsec respectively.
The central \dfourk\ excess averaged over 100 times matching is recorded for every satellite and we do this analysis separately for satellites in low ($10^{-11.4}-10^{-10.4}\,\mathrm{yr}^{-1}$; the upper row of Fig. \ref{fig:psd}) and high ($10^{-10.4}-10^{-9.4}\,\mathrm{yr}^{-1}$; the bottom row of Fig. \ref{fig:psd}) sSFR ranges.
The right column shows the locally averaged results using the locally weighted regression method LOESS by \citet{Cleveland1988} as implemented\footnote{We use the Python package \textsc{loess} v2.0.11 available from https://pypi.org/project/loess/} by \citet{2013MNRAS.432.1862C}, to reveal the underlying trend.
We adopt a smoothing factor \texttt{frac} = 0.3, and a linear local approximation, but the conclusion does not depend on these certain parameter choices.

In the upper right panel of Fig. \ref{fig:psd} for satellites of low sSFR, LOESS reveals certain structure of \dfourk\ excess at low groupcentric radii.
The largest \dfourk\ excess is not seen evenly for all the galaxy populations near cluster center, but is particularly linked with the satellites of either small or large relative velocities.
Satellites of intermediate velocities of about $|\Delta v|/\sigma_{200} = 0.7$ only show moderate \dfourk\ excess comparable to those at much larger groupcentric radii.
This result indicates an apparent connection between the \dfourk\ excess and the orbit configuration of satellites.
In the lower right panel for satellites of high sSFR, which are probably in the early stages of environmental processing, the \dfourk\ excess is low but noteworthily shows the same pattern as the low-sSFR satellites.
The consistency suggests that the observed pattern of locally averaged \dfourk\ excess reflects the true trend underlying the noisy data in the left column.

\section{Summary and discussion}
\label{sec:discuss}

In this paper, we have investigated the environmental dependence of the relative difference in sSFR radial gradient for 0.1 million SDSS galaxies at $z \sim 0$.
We compare the central sSFR, indicated by indices \dfourk\ and \hda\ measured from SDSS fiber spectra, between satellite and isolated galaxies at the same total sSFR, so that we extract how galaxy environment affects the sSFR radial gradient in a relative sense.
With fiber coverage properly matched for the comparison, the large sample size facilitates the study of detailed correlations with a variety of environmental properties when the mass and star formation level of galaxies are controlled.
Our findings are summarized as below:
\begin{enumerate}[(i)]
  \item Low-mass satellite galaxies ($\mathcal{M}_{\star}=10^9-10^{9.8}\,\mathcal{M}_{\odot}$) below the SFMS have lower central sSFR compared to isolated counterpart galaxies at given total sSFR (Fig. \ref{fig:100bs1}).
  \item The phenomenon of more suppressed central star formation (i.e. the central \dfourk\ excess at given total sSFR) among low-mass satellites becomes more noticeable in host halos of higher mass (equivalently of more member galaxies), and when closer to the group center, while more massive galaxies below the SFMS show consistent trend but with smaller amplitude (Fig. \ref{fig:Edep}).
  The dependence on halo mass is of first-order importance and the dependence on groupcentric radius is secondary (Fig. \ref{fig:Edep1}).
  \item In the center of massive halos, phase-space diagram reveals that the phenomenon is strongest among satellites of either lowest or highest relative velocities to the halo (Fig. \ref{fig:psd}), indicating the connection between the suppressed central star formation and orbital configuration of satellite galaxies.
\end{enumerate}

\begin{figure}
    \centering
    \includegraphics[width=\linewidth]{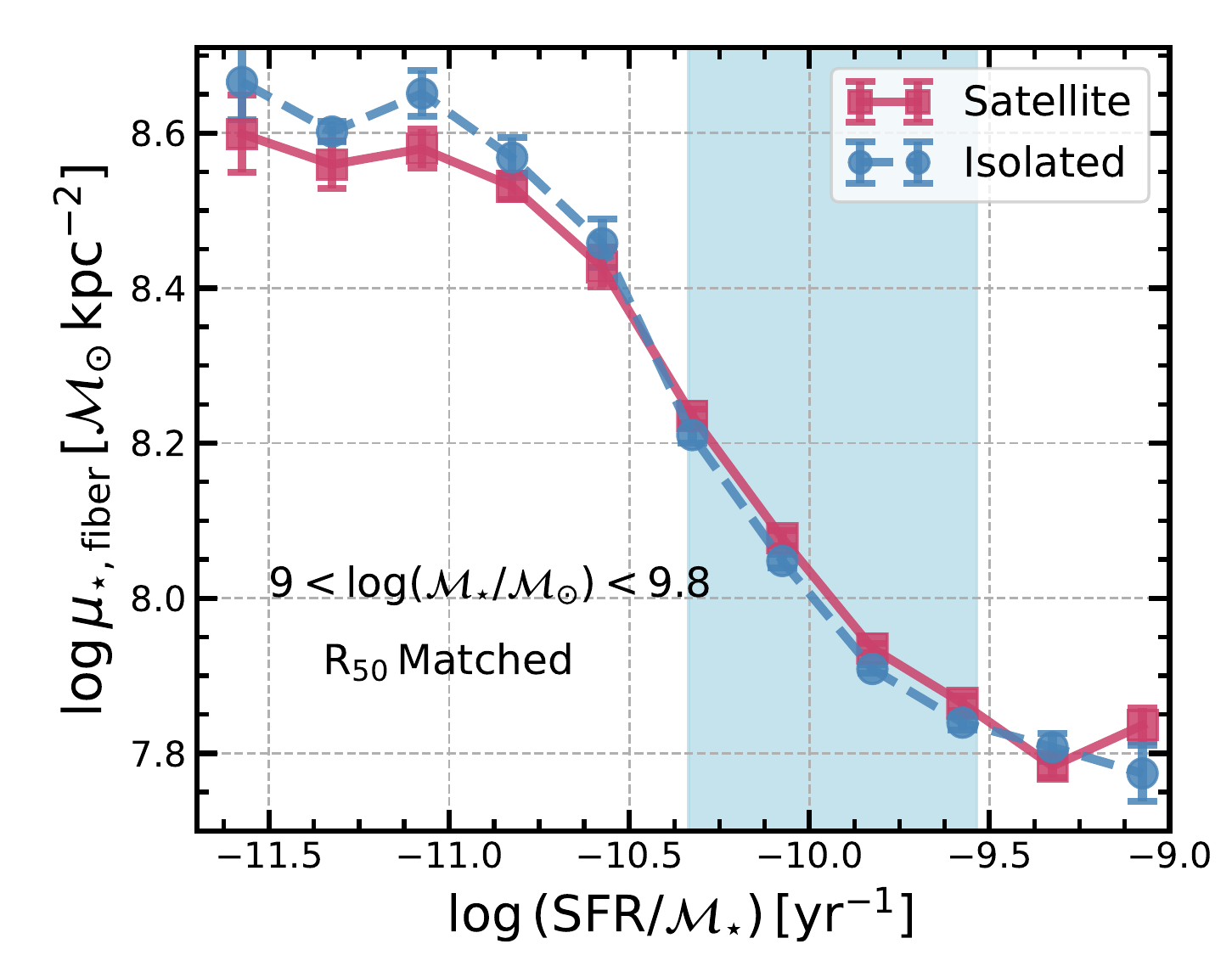}
    \caption{The stellar mass surface density within the fiber aperture as a function of total sSFR for low-mass satellite and isolated galaxies.}
    \label{fig:fiber_mu}
\end{figure}

\begin{figure*}
	\begin{center}
		\includegraphics[width=0.9\textwidth]{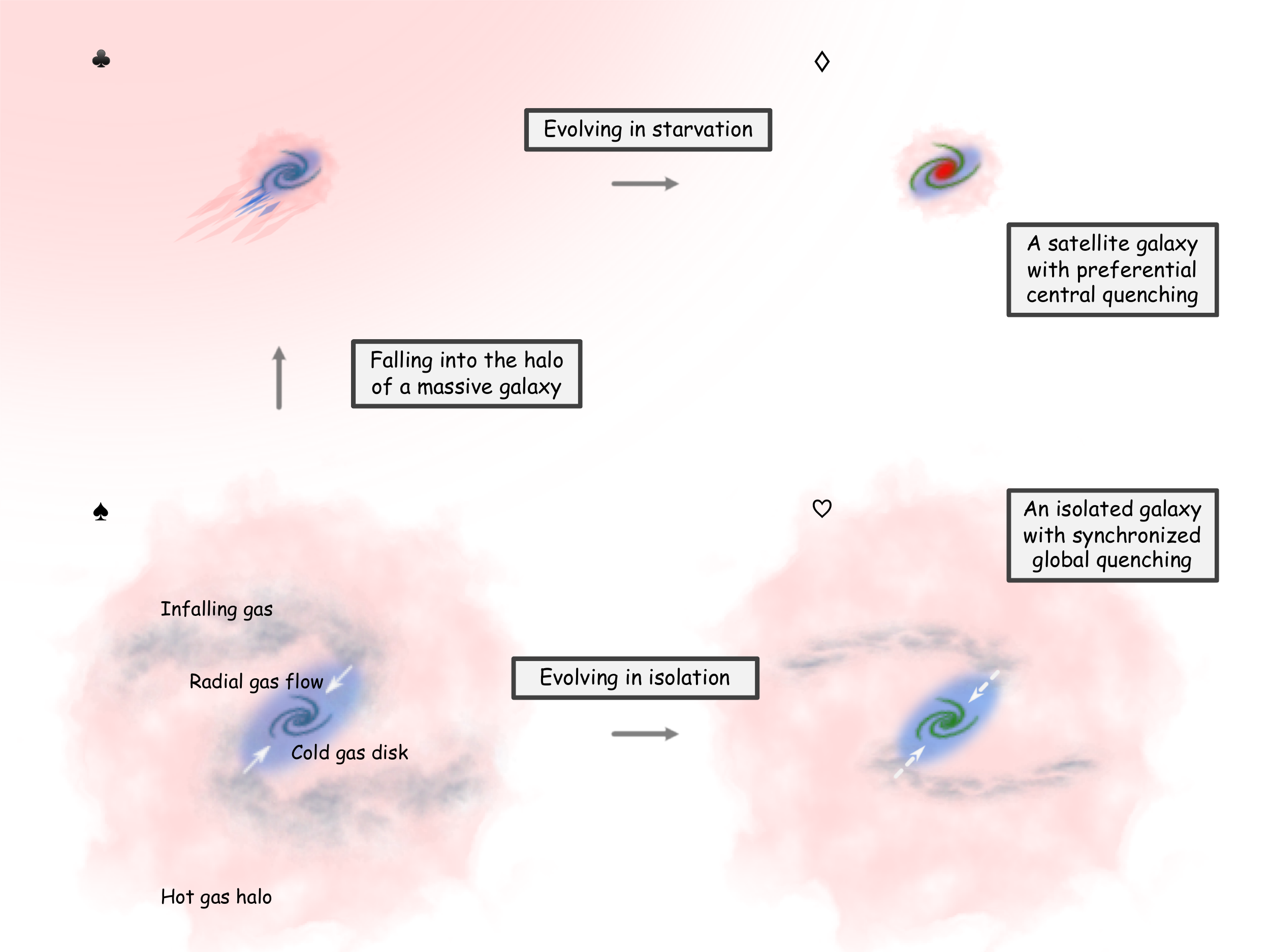}
		\caption{
    A schematic illustration of the proposed scenario explaining how gas stripping in massive halos can render the quenching of satellites more inside-out.
    An actively star-forming galaxy in isolation (panel $\spadesuit$) has an extended cold gas disk and a more extended hot gas halo.
    Cold gas disk is replenished by infalling gas cooled out of hot gas which also drives significant gas radial flow (denoted as white arrows) on the disk due to mismatch of angular momentum.
    If such a galaxy falls into the hot gas halo of another much more massive galaxy and becomes a satellite ($\spadesuit \Rightarrow \clubsuit$) both its hot gas halo and the outskirt of its cold gas disk can be largely stripped during orbiting motion after which gas infalling and gas radial inflowing stop.
    In such starvation the central part of this galaxy quenches first due to the high star formation efficiency there (panel $\diamondsuit$).
    By contrast if this galaxy keeps evolving in isolation ($\spadesuit \Rightarrow \heartsuit$), the diminishing gas cooling and infalling (expected for quenching) do not terminate gas radial inflow immediately.
    Thus the central part can still be replenished during the overall quenching and so the quenching can be radially synchronized.
		}
		\label{fig:illus}
	\end{center}
\end{figure*}

\subsection{The physical mechanisms}\label{subsec:phy}

The more suppressed central star formation of satellites compared to field galaxies of the same total sSFR suggests that additional physical processes in galaxy groups make the quenching of star formation happen more inside-out.
The environmentally promoted inside-out quenching is especially shown by the sharp increase of central \dfourk\ with decreasing total sSFR among the low-mass satellites (Fig. \ref{fig:100bs1}).
The SFR profiles of low-mass satellites can even deviate more from the profiles of their field counterparts because we find, as shown in Fig. \ref{fig:fiber_mu}, that the central stellar mass density within fiber area of low-mass satellites of low sSFR is smaller than field galaxies which is consistent with \citealt{2017MNRAS.464.1077W}.
The stellar mass measurements inside fiber area are taken from the MPA-JHU catalogue, with a small mean difference of $\sim0.1$ dex compared to GSWLC stellar mass \citep{2016ApJS..227....2S}.
The lower central stellar mass density of satellites seems to result from the integrated effect of their suppressed central star formation.

So far it is unclear, among miscellaneous physical processes occurring in group environment, which mechanism is mainly responsible for the central \dfourk\ excess of low-mass satellite galaxies.
In Fig. \ref{fig:Edep1}, we see that the high \dfourk\ excess is preferentially found in massive clusters, especially in the cluster center.
The strongest effect in the cluster center is seen among satellites with either lowest or highest velocities on the phase-space diagram.
The former satellite population with lowest velocity generally have low orbital energy as a result of their low potential energy (i.e. at the bottom of potential well) and low kinetic energy.
Suggested by simulations \citep[e.g.,][]{2013MNRAS.431.2307O}, these satellites joined the cluster during ancient infalls and have thus been trapped in the center for long time.
The latter satellite population with high velocity in the vicinity of cluster center are suggested to be recent infallers that are experiencing their first or second pericenter.
Projection of velocity and position of satellites can smear such connection between orbital properties and the position on phase-space diagram.
However the clear consistency across satellite populations of high and low sSFR living in a large number of different groups rejects the possibility that the result is due to random projection.
From the perspective of environmental effect, the former satellite population experience in long term the enormous tidal force from the massive cluster, which anti-scales with cubic groupcentric distance and can play an important role in shaping the star formation and morphology of galaxies \citep{1984ApJ...276...26M,1990ApJ...350...89B}.
The latter satellite population, when they pass the orbit pericenter, on short timescales not only do they feel the strong cluster tidal field but also large ram pressure due to both the high density of intracluster medium and their high velocities.
The middle panel of Fig. \ref{fig:Edep1} shows that there is non-negligible \dfourk\ excess at even the outskirt of massive halos, where the cluster tidal field weakens dramatically.
While hydrodynamic gas stripping can still be effective in the outskirt of halos for satellites with high velocities, and some cases were indeed caught in action \citep[e.g.,][]{2018MNRAS.476.4753J}.
This also coincides with the fact shown in the upper right panel of Fig. \ref{fig:psd} where we see that at large groupcentric radii satellites of higher velocities manifest larger central \dfourk\ excess.
These together seem to suggest that both tidal and hydrodynamic interactions are responsible for the phenomenon of suppressed central star formation of satellite galaxies.

\begin{figure*}
	\begin{center}
		\includegraphics[width=0.96\textwidth]{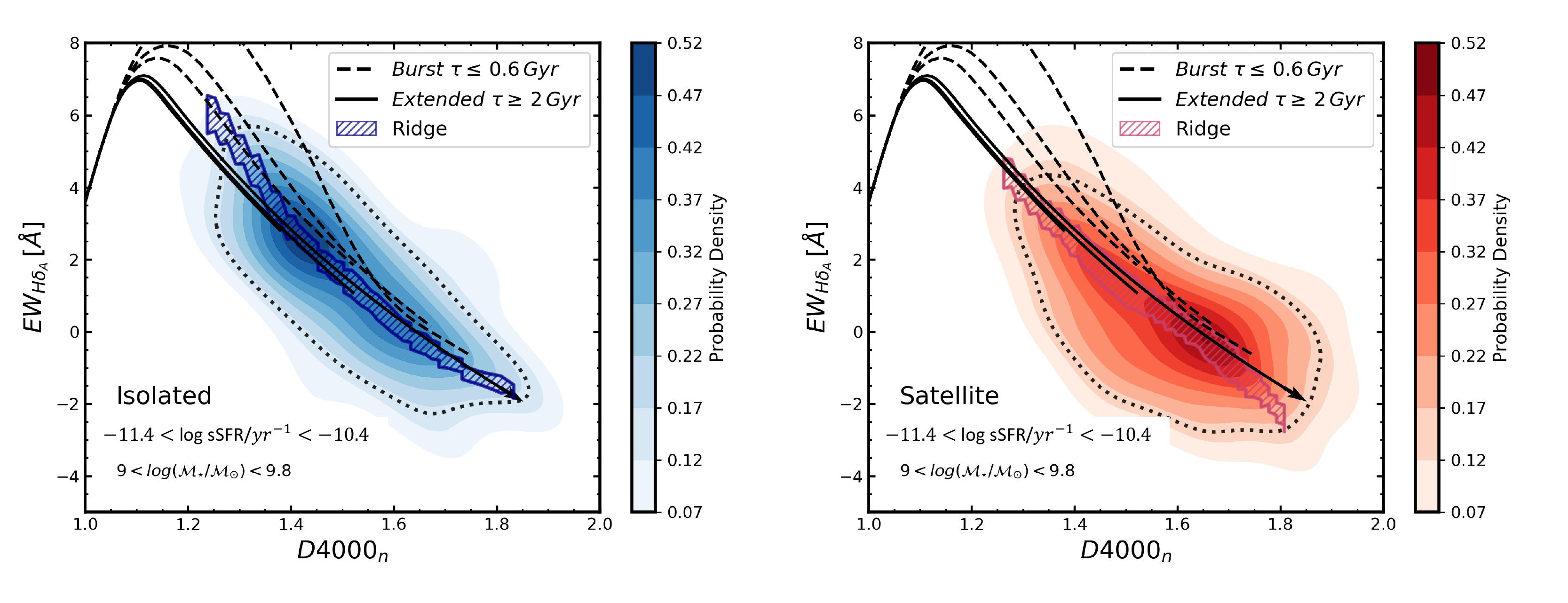}
		\caption{
      Comparison between the probability density functions (filled contours) of low-mass satellite (right panel) and isolated (left panel) galaxies in sSFR range $10^{-11.4}-10^{-10.4}\,\mathrm{yr}^{-1}$ on the \hda-\dfourk\ plane with evolutionary tracks (black lines) generated from BC03 models.
      The probability density functions are derived via kernel density estimation with $V_{\mathrm{max}}$ corrections.
      The contour enclosing 68\% of total probability is shown as a dotted line.
      The ridges (shaded regions) of density distributions are identified, following \citealt{2016ApJ...823...18C}, as the most prominent track for each galaxy population.
      The uncertainty of ridges is assessed by 1,000 bootstrap samples and one sigma confidence intervals are shown.
      Model tracks are generated by convolving model spectra with exponentially declining SFHs ($\mathrm{SFR}\,\propto\,\mathrm{exp}(-t/\tau)$) of short (dashed lines; from up to down, $\tau=0.2,0.4,0.6\,\mathrm{Gyr}$) and long (solid lines; from up to down, $\tau=2,4,6\,\mathrm{Gyr}$) characteristic timescales.
      See more details of model tracks in the text.
		}
		\label{fig:sfhs1}
	\end{center}
\end{figure*}

It is known that tidal interactions can strip the loosely bound peripheral gas of galaxies in synergy with the hydrodynamic gas stripping, which together result in galaxy starvation and prevent further gas accretion \citep{2002ApJ...577..651B}.
In starvation, galaxies tend to quench inside-out due to the one order of magnitude faster gas depletion in the center than in the outer part \citep{2008AJ....136.2782L}.
Starvation promotes inside-out quenching also because that the radial gas inflows on galactic disks may be largely reduced.
As accretions of gas from gaseous halos can drive radial gas inflows due to even just a small mismatch of angular momentum between the accreted gas and the disks \citep{2016MNRAS.455.2308P}.
\citealt{2012MNRAS.426.2266B} reports that this process is one of the most dominant processes inducing radial inflows, making the process an important channel of fuelling central star formation.
So the central star formation is less supported in a satellite with a largely stripped gaseous halo (i.e. in starvation).
By contrast, during the quenching of isolated galaxies, as long as the hot gaseous halos still exist, their central parts are more likely to be fed by cold gas compared to those highly stripped satellites.
We illustrate this scenario in Fig. \ref{fig:illus} ($\spadesuit - \clubsuit - \diamondsuit$ for satellites and $\spadesuit - \heartsuit$ for isolated galaxies).

Starvation as an explanation for the phenomenon shown in this work seems to be in line with \citealt{2015Natur.521..192P, 2019MNRAS.tmp.2878T} which point out the major role of starvation in quenching the low-mass galaxy populations and the growing significance of starvation in denser environments.
Though not reporting on the spatial distribution of star formation, \citealt{2017MNRAS.464..508B} found that the same mechanism drives the enhancement of gas metallicity of satellite galaxies in the EAGLE simulations \citep{2015MNRAS.446..521S}.
They found that the central gas metallicity is enhanced effectively when starvation suppresses the radial inflow of gas, which is predominantly metal-poor.

\subsubsection{Evidence in recent star formation history}\label{subsubsec:sfhs}

The scenario above can have detectable consequences for the recent star formation history (SFH) in the central part of satellite and isolated galaxies.
We probe the recent SFH by the combination of \dfourk\ and \hda\ which trace stellar populations of different ages (see also \citealt{2003MNRAS.341...33K}).
Fig. \ref{fig:sfhs1} shows satellite and isolated galaxies of low mass and low sSFR on the \hda-\dfourk\ plane, overlaid with evolutionary tracks of \citealt{2003MNRAS.344.1000B} models (BC03).

The probability density function of galaxies (filled contours) is derived via kernel density estimation with $V_{\mathrm{max}}$ corrections.
We use Gaussian kernel of width determined by Scott's rule \citep{Scott2015Multivariate}.
Then, we identify the ridge line (following \citealt{2016ApJ...823...18C} and is shown by hatched area) for each density distribution as the representative track for the galaxy population.
In producing model tracks of exponentially declining SFHs (black dashed lines: declining timescale $\tau=0.2,0.4,0.6\,\mathrm{Gyr}$; black solid lines: $\tau=2,4,6\,\mathrm{Gyr}$), we use MILES stellar library of solar metallicity and Padova 1994 library for stellar evolution prescription.
Using other empirical or theoretical stellar libraries and other stellar evolution prescriptions provided in BC03 generates model tracks significantly incompatible with our data.

The contours show that, compared to isolated galaxies (left panel), a significantly higher fraction of satellites (right panel) populate the lower right area indicating again the suppressed central star formation of satellite galaxies.
Moreover, the distribution of isolated galaxies is more concentrated around the ridge while that of satellites has a broader shape.
This may imply that group environment can diversify the SFH of galaxies.
Noteworthily, while the ridge line of satellites can be overall matched by continuously declining SFHs of long timescales over Gyrs, the ridge line of isolated galaxies deviates obviously toward models of shorter timescales.
Such deviation is due to a non-negligible fraction of isolated galaxies with high \hda\ at given \dfourk.
As \hda\ mainly traces A-type young stars, this elevated \hda\ indicates the significance of recent burst of star formation (see also the Fig. 6 in \citealt{2003MNRAS.341...33K}) in the central part of isolated galaxies.

The observed difference in recent SFH between satellite and isolated galaxies fits into the scenario described before.
The existing hot gas halo of low-sSFR isolated galaxies can still fuel some small bursts of star formation, when the inefficient gas cooling (expected from low sSFR) is only able to drive gas radial flows episodically.
By contrast, the central part of satellites in starvation are more likely to turn red quiescently and smoothly when without further gas supply.

\subsection{Comparison with previous works}\label{subsec:comparison}

The discussion above does not incorporate gas stripping caused outside-in quenching as a major driver of the cessation of total star formation in group environments.
Instead, environments are observed to render quenching of low-mass galaxies more inside-out.
However, it has to be clarified that the results do not indicate that gas stripping does not influence outer star formation to any extent.
The results only suggest that the inner parts of galaxies contribute primarily to the total decline of star formation under environmental effects, while the suppression of star formation in the outskirt is only secondary.
The conclusion is echoed by \citealt{2019ApJ...872...50L}, who found that inside-out quenching is the highly dominant channel even for satellites in massive halos and the fraction of galaxies experiencing outside-in quenching does not depend on halo mass at all.

The same conclusion was not reached by many other works in the literature, which are also contradicting among themselves.
Using 1,494 MaNGA galaxies, \citealt{2018MNRAS.476..580S} compared the sSFR radial profiles of central and satellite galaxies.
Their Fig. 7 indicates that, in the intermediate and high mass bins, the sSFR of satellites are systematically lower than the central galaxies particularly outside 0.5 effective radius.
For galaxies in the low-mass bin, this pattern appears to be reversed, showing more inside-out quenching for satellites.
In spite of the general consistency among low-mass galaxies between \citealt{2018MNRAS.476..580S} and our work, our data do not indicate the outside-in quenching for massive satellite galaxies.
\citealt{2019A&A...621A..98C} used a smaller sample of 275 late-type CALIFA galaxies and carried out similar analyses.
As entirely opposed to the results in \citealt{2018MNRAS.476..580S}, for low-mass galaxies in groups they found more suppressed star formation in the outer parts compared with galaxies in the field, and for the massive galaxies, more suppressed in the inner parts.
Rather than being suppressed, the low-mass satellite galaxies studied by \citealt{2019MNRAS.489.1436L} show centrally enhanced star formation in the densest environments.
Apart from these recent works based on IFS data, \citealt{2009MNRAS.394.1213W} studied the g-r colour profiles of galaxies in the SDSS Data Release 4.
They found outside-in quenching pattern for the satellite galaxies in their high mass bin.
In their low-mass bin, the colour profiles of the satellites are globally redder compared to the central galaxies.
Their sample almost does not cover the low-mass range of our data.

The intricate discrepancies between works in the literature can result from a variety of reasons.
Noteworthily, the samples were selected with diverse criteria.
For example, \citealt{2017MNRAS.464..121S} only selected galaxies with central regions classified as star-forming by emission line diagnostics.
This may have biased their sample against centrally quenched galaxies, which would have weak emission lines in the center.
\citealt{2019MNRAS.489.1436L} introduced thresholds for signal to noise of emission lines during the sample selection.
The sample of \citealt{2019A&A...621A..98C} was preselected by Hubble type.
Moreover, a problem in some previous studies is that sSFR radial profiles are not compared at the same level of total sSFR for galaxies in different environments.
While many IFS studies \citep[e.g.,][]{2018MNRAS.477.3014B,2018ApJ...856..137W} have shown that sSFR radial gradients clearly depend on the level of total sSFR.
Therefore, extracting a more unambiguous dependence on environment needs better control of total sSFR, as we have done in this work.

\section*{Acknowledgements}
BW acknowledges the elaborated and constructive comments from the anonymous referee which significantly helped improve this manuscript.
BW thanks Li Shao for his insightful and decisive comments on this work, and thanks Jing Wang, Min Du, and Jingjing Shi for the fruitful discussions with them.

Funding for the SDSS and SDSS-II has been provided by the Alfred P. Sloan Foundation, the Participating Institutions, the National Science Foundation, the U.S. Department of Energy, the National Aeronautics and Space Administration, the Japanese Monbukagakusho, the Max Planck Society, and the Higher Education Funding Council for England.
The SDSS is managed by the Astrophysical Research Consortium for the Participating Institutions. The Participating Institutions are the American Museum of Natural History, Astrophysical Institute Potsdam, University of Basel, University of Cambridge, Case Western Reserve University, University of Chicago, Drexel University, Fermilab, the Institute for Advanced Study, the Japan Participation Group, Johns Hopkins University, the Joint Institute for Nuclear Astrophysics, the Kavli Institute for Particle Astrophysics and Cosmology, the Korean Scientist Group, the Chinese Academy of Sciences (LAMOST), Los Alamos National Laboratory, the Max-Planck-Institute for Astronomy (MPIA), the Max-Planck-Institute for Astrophysics (MPA), New Mexico State University, Ohio State University, University of Pittsburgh, University of Portsmouth, Princeton University, the United States Naval Observatory, and the University of Washington.

\section*{Data Availability}
The data used in this work are all publicly available.
We take the MPA-JHU catalogue from https://wwwmpa.mpa-garching.mpg.de/SDSS/DR7/ and the GSWLC catalogue from https://salims.pages.iu.edu/gswlc/ and the group catalogue from https://gax.sjtu.edu.cn/data/Group.html for SDSS galaxies.

%%%%%%%%%%%%%%%%%%%%%%%%%%%%%%%%%%%%%%%%%%%%%%%%%%



\bibliographystyle{mnras}

%%%%%%%%%%%%%%%%%%%%%%%%%%%%%%%%%%%%%%%%%%%%%%%%%%

\newpage

% %%%%%%%%%%%%%%%%% APPENDICES %%%%%%%%%%%%%%%%%%%%%

\appendix

\section{Feasibility of probing sSFR radial gradient by central and total sSFR}\label{app:fea}

% %%%%%%%%%%%%%%%%%%%%%%%%%%%%%%%%%%%%%%%%%%%%%%%%%%%%%%%%%%%%%%%%%%%%%%%%%%%%%%

\begin{figure}
	\includegraphics[width=0.47\textwidth]{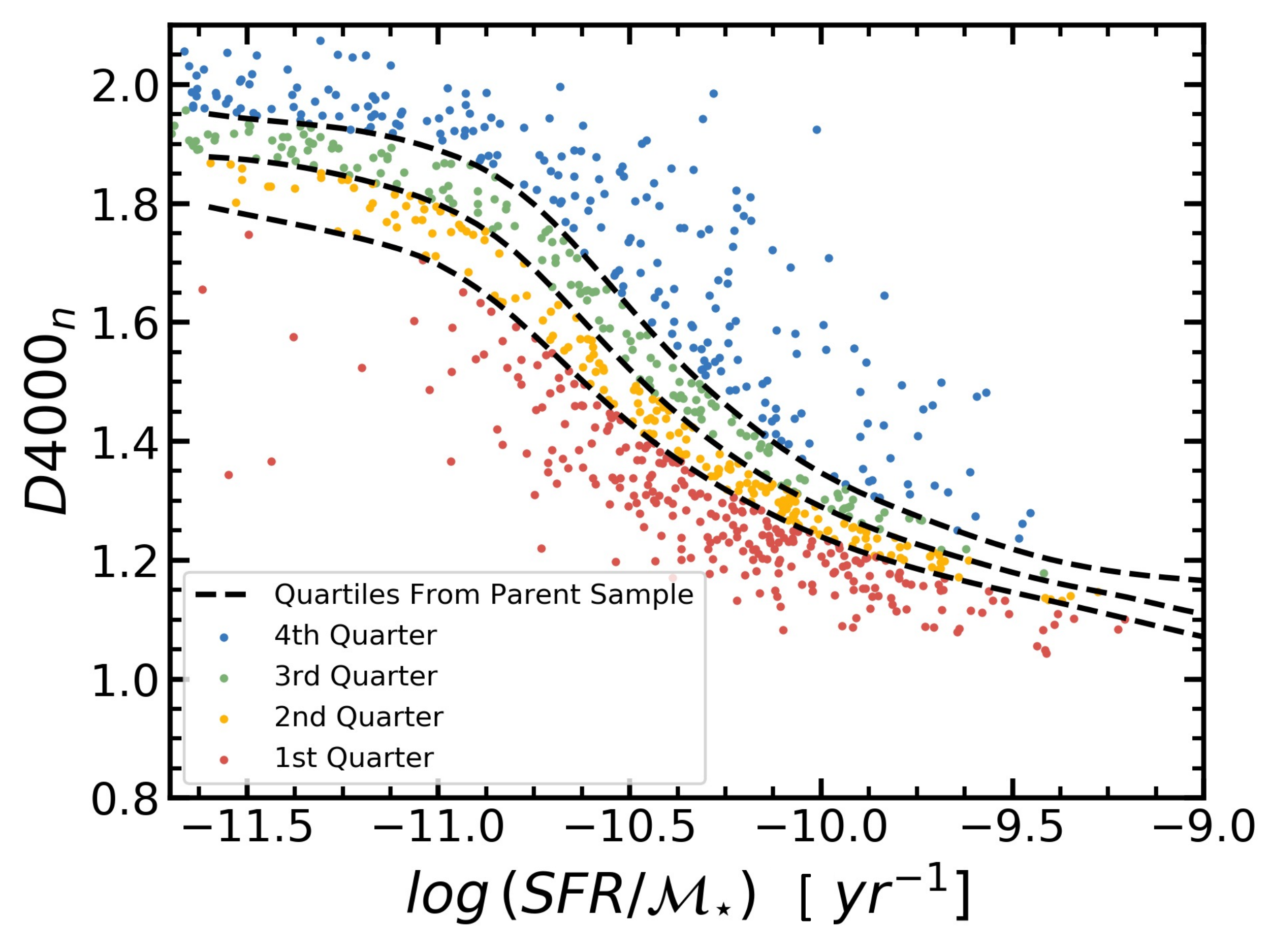}
	\caption{Galaxies with available IFS data in P{\sc ipe}3D DR14, on central \dfourk\ vs. total sSFR plane.
  Black dashed lines are quartiles defined by the distribution of \dfourk\ at given total sSFR of the parent sample.
		}
	\label{fig:s3}
\end{figure}

% %%%%%%%%%%%%%%%%%%%%%%%%%%%%%%%%%%%%%%%%%%%%%%%%%%%%%%%%%%%%%%%%%%%%%%%%%%%%%%

\begin{figure*}
	\begin{center}
		\includegraphics[width=0.45\textwidth]{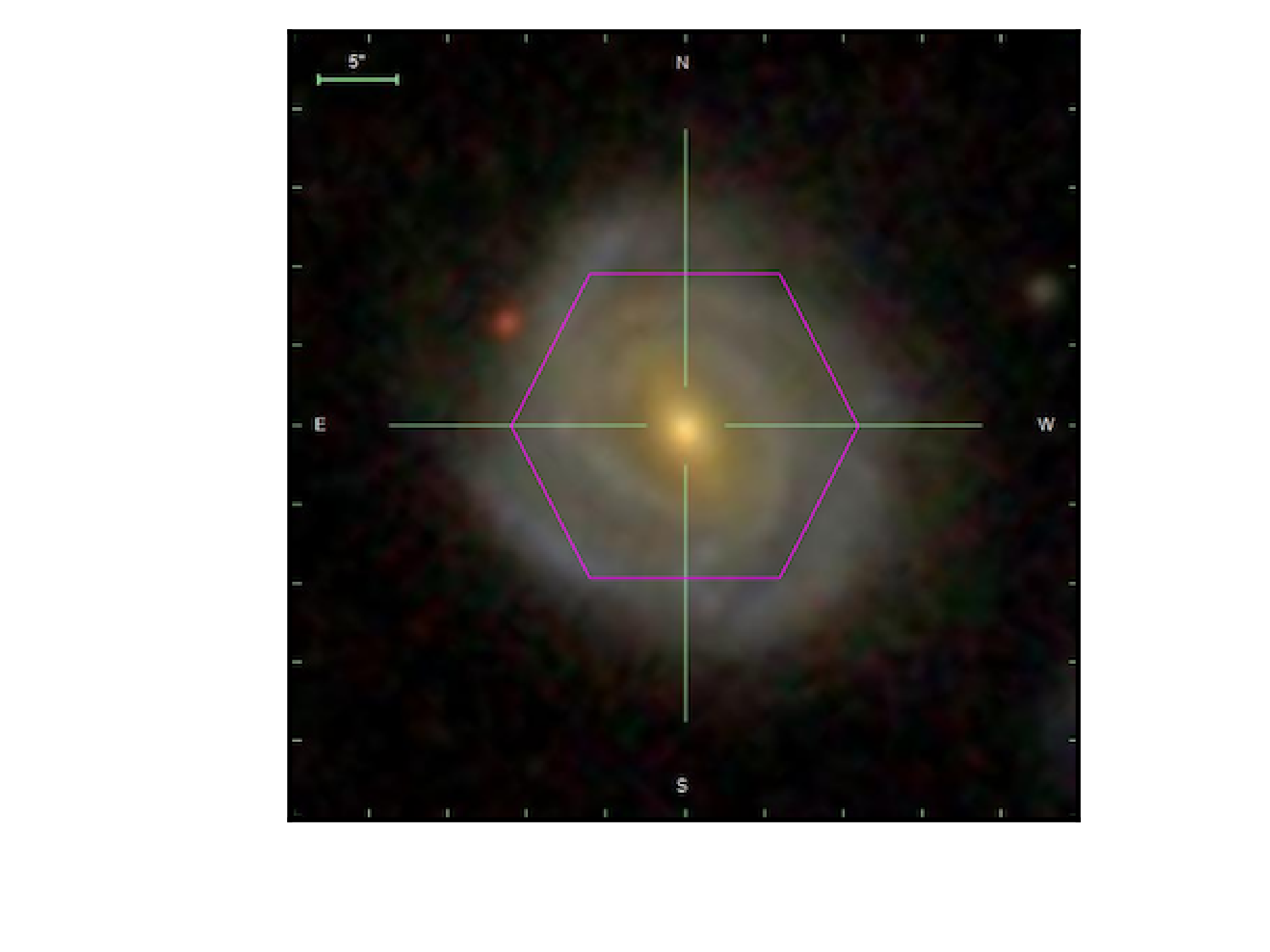}
    \includegraphics[width=0.45\textwidth]{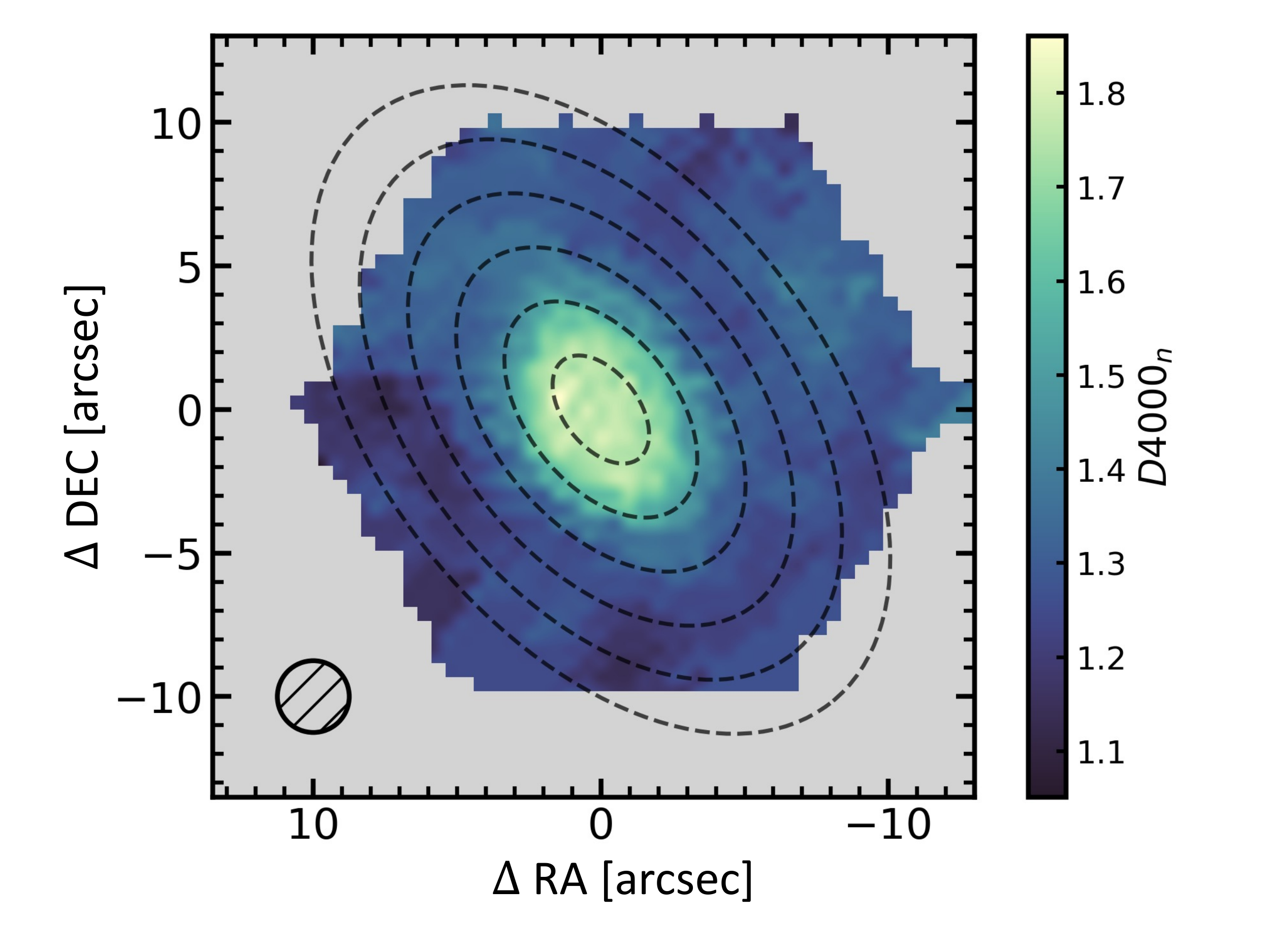}
		\caption{An example showing the \dfourk\ map and its radial binning of a galaxy with MaNGA observation (MaNGA ID: 8979-6102).
    Left: g--r--i SDSS image composite of the galaxy with overlaid MaNGA hexagonal FoV.
    Right: \dfourk\ map of the galaxy and radial binning (black dashed lines) described in the text.
    The MaNGA PSF is shown as a hatched circle in the bottom-left corner.
			}
		\label{fig:s31}
	\end{center}
\end{figure*}

% %%%%%%%%%%%%%%%%%%%%%%%%%%%%%%%%%%%%%%%%%%%%%%%%%%%%%%%%%%%%%%%%%%%%%%%%%%%%%%

\begin{figure*}
	\begin{center}
		\includegraphics[width=0.8\textwidth]{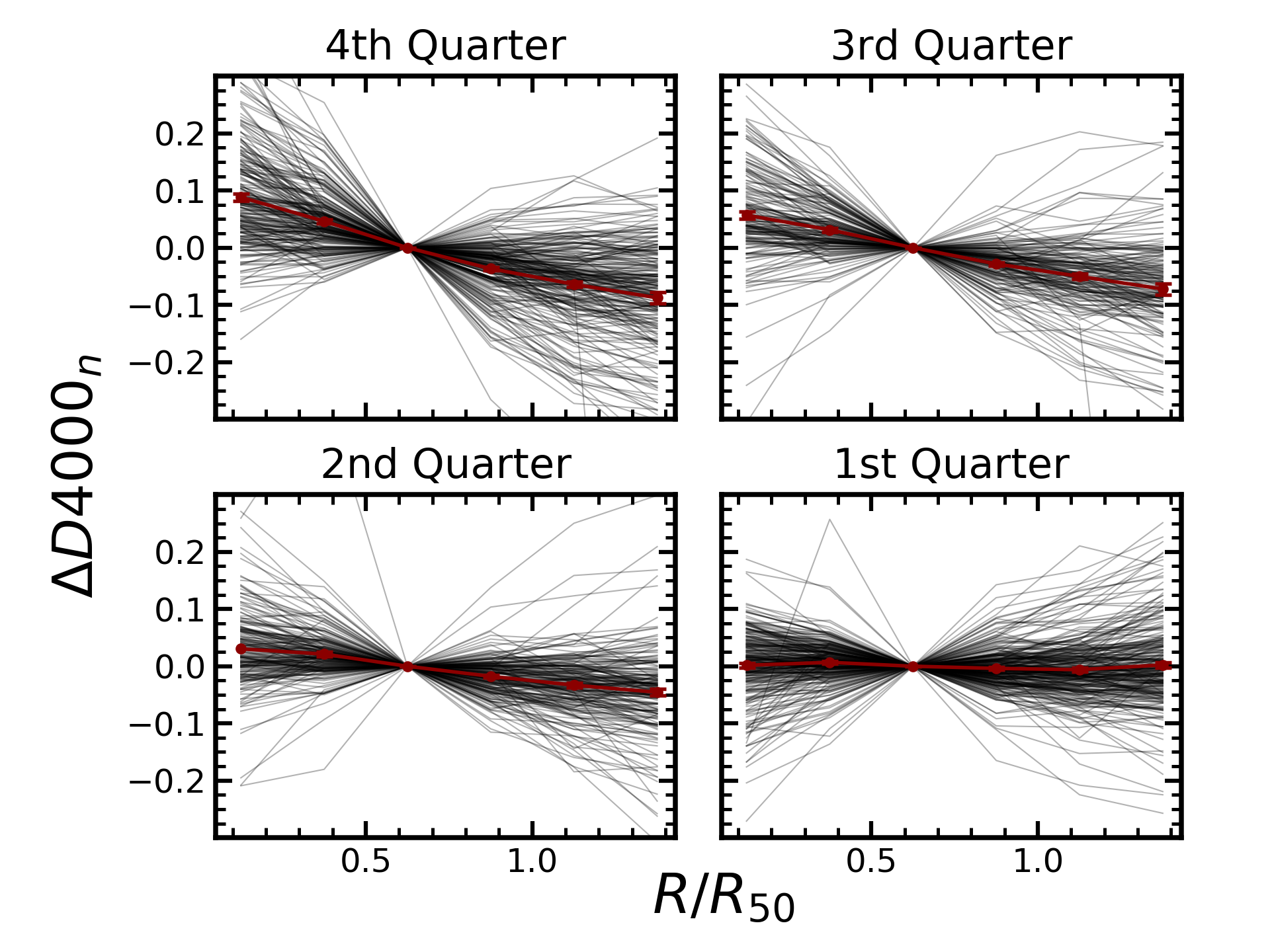}
		\caption{\dfourk\ radial profiles (black lines) of galaxies in different quarters, each normalized to the third radial bin.
    Red lines indicate the median relations with one sigma errors estimated through 1000 bootstrap samples.
			}
		\label{fig:s33}
	\end{center}
\end{figure*}

% %%%%%%%%%%%%%%%%%%%%%%%%%%%%%%%%%%%%%%%%%%%%%%%%%%%%%%%%%%%%%%%%%%%%%%%%%%%

The basic idea of our analysis is that at a given total sSFR, the variation of central sSFR reflects the change in sSFR radial gradient across the disks.
Below we prove this in a statistical sense by using a small sample of galaxies with MaNGA IFS data.

First, at given total sSFR we divide our sample into four quarters with respect to their central \dfourk\ values and the loci of the quartiles are shown as black dashed lines in Fig. \ref{fig:s3}.
By doing this we roughly classify our sample into four subsamples with different sSFR radial gradients.
For galaxies in each quarter we search for reduced MaNGA cubes in a value-added catalogue P{\sc ipe}3D \citep{2016RMxAA..52...21S,2016RMxAA..52..171S}, released as a part of SDSS DR14.
This allows us to look into their spatially resolved maps of sSFR.
We find 262, 187, 166 and 230 galaxies included in P{\sc ipe}3D for the first to the fourth quarters (red to blue points in Fig. \ref{fig:s3}) respectively after excluding cubes flagged as bad.

We derive the \dfourk\ radial profile of these galaxies as follows.
Their \dfourk\ maps are binned, with a step of 0.25 $\mathrm{R_{50}}$, by six ellipses of position angle and ellipticity determined for the galaxy by P{\sc ipe}3D pipeline \citep{2016RMxAA..52..171S}.
An example of the \dfourk\ map and the binning for a galaxy belonging to the forth quarter are shown in the right panel of Fig. \ref{fig:s31} together with the corresponding SDSS g--r--i image composite in the left panel.
The expected significantly negative \dfourk\ radial gradient (because the galaxy is in the 4th quarter) is clearly seen on that map.
We then measure \dfourk\ radial profiles by calculating the median \dfourk\ of spaxels in each radial bin without cut of signal-to-noise ratio.
The derived profiles for galaxies in different quarters, each normalized to the third radial bin, are displayed in Fig. \ref{fig:s33} as black lines.
In each quarter, the median relation is shown by the red line with one sigma error estimated by 1000 bootstrap samples.
Fig. \ref{fig:s33} shows a systematic change of \dfourk\ radial gradient as expected from our original classification according to the central and total sSFR of galaxies, and proves the feasibility of our method.
We get the same conclusion by analyzing the maps of \hda\ and \halpha.

\section{Confirming the main result using NUV+MIR SFRs}\label{app:nuvmir}

\begin{figure*}
	\begin{center}
		\includegraphics[width=0.9\textwidth]{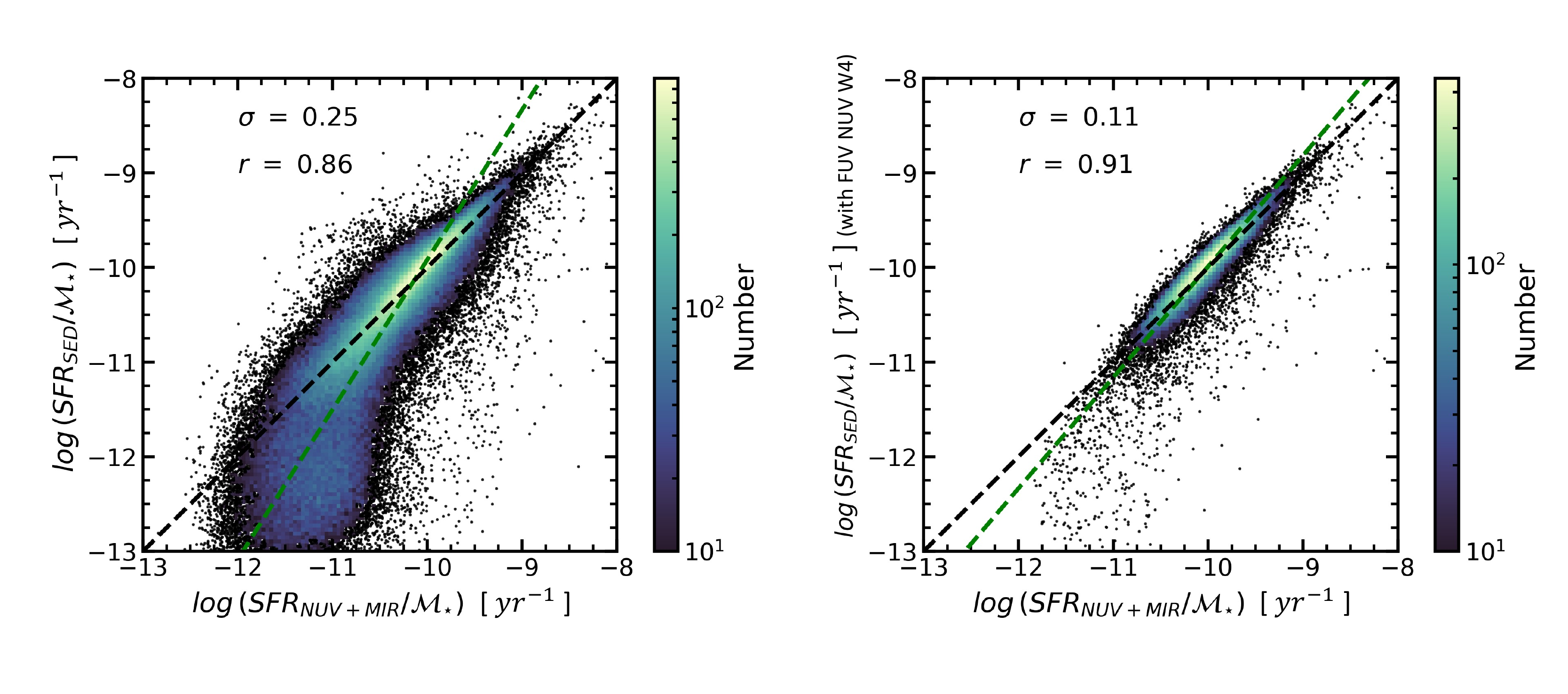}
		\caption{Left: comparison between SFRs in GSWLC-M2 derived from UV to MIR SED fitting and SFRs derived from GALEX NUV plus WISE W4 luminosities, both divided by GSWLC-M2 stellar mass.
    Black dashed line shows a 1:1 relation and green dashed line presents the best fitting linear relation using orthogonal distance regression.
    Scatter from the best fitting linear relation ($\sigma$) and Pearson linear correlation coefficient ($r$) are shown on the upper part.
    Right: The same as the left panel but limited to galaxies with all GALEX FUV and NUV and WISE W4 detections.
			}
		\label{fig:s1}
	\end{center}
\end{figure*}

%%%%%%%%%%%%%%%%%%%%%%%%%%%%%%%%%%%%%%%%%%%%%%%%%%%%%%%%%%%%%%%%%%%%%%%%%%%%%%

In this appendix we reproduce the left panel of Fig. \ref{fig:100bs1} using SFRs measured directly from GALEX NUV and WISE W4 fluxes.

For galaxies in GSWLC-M2 catalogue, we retrieve their GALEX NUV magnitude from the GALEX final data release (DR6/7)
\footnote{http://galex.stsci.edu/GR6/}
via MAST Casjobs
\footnote{https://galex.stsci.edu/casjobs/}.
The nontrivial matching between SDSS and GALEX has been done in \citealt{2016ApJS..227....2S} so that we can retrieve GALEX data by directly using GALEX \textit{objid} in Casjobs.
Among 305,094 galaxies in GSWLC-M2 that have a valid GALEX \textit{objid}, 280,101 galaxies are detected in NUV after excluding several hundred duplications.
The WISE four bands photometry (W1 to W4 at 3.6, 4.6, 12 and 22 $\mu m$ respectively) is taken from unWISE
\footnote{http://unwise.me}
\citep{2016AJ....151...36L}, where SDSS detections served as forced photometry priors in the reduction of WISE data.
Since the unWISE photometry is based on SDSS DR10 detections, every source has already been matched with SDSS.
We directly take the unWISE data for every galaxy in GSWLC-M2 through SDSS \textit{ObjID}.

Before converting GALEX NUV magnitude to luminosity
\footnote{https://asd.gsfc.nasa.gov/archive/galex/FAQ/counts\_background.html}
, galactic reddening is corrected for using color excess derived from dust maps of \citealt{1998ApJ...500..525S} assuming a \citealt{1989ApJ...345..245C} extinction curve.
And WISE W4 flux is corrected for ``red" sources by giving a 8\% reduction
\footnote{http://wise2.ipac.caltech.edu/docs/release/allsky/expsup/sec4\_4h.html\#example}
.
With all data requirements included in above procedures, finally we got 203,191 galaxies (56.2\% of GSWLC-M2) having corrected NUV and W4 luminosities.

We take the NUV star formation calibration in \citealt{2013seg..book..419C}, which is obtained from stellar population models assuming a Kroupa IMF in mass mange $0.1-100\,\mathcal{M}_{\odot}$ and constant star formation rate over $100\,\mathrm{Myr}$.
We further applies a factor of 95\% to adjust from Kroupa to Chabrier \citep{2007ApJS..173..267S} as GSWLC SFRs are based on the latter and finally:

\begin{equation}\label{nuvsfr}
\mathrm{SFR}_{\mathrm{NUV}}/(\mathcal{M}_{\odot}\,yr^{-1})=6.46\times10^{-44}\times L_{\mathrm{NUV},\,\mathrm{total}}/(\mathrm{erg}\,\,s^{-1})
\end{equation}

The NUV flux received by GALEX is the unobscured part, we compensate for dust extinction according to \citealt{2011ApJ...741..124H}:

\begin{equation}\label{dust}
L_{\mathrm{NUV},\,\mathrm{total}}=L_{\mathrm{NUV},\,\mathrm{obscured}}+2.26\times L_{25\,\mu m}
\end{equation}
where $L_{25\,\mu m}$ stands for luminosity in Infrared Astronomical Satellite (IRAS) band centered at 25 $\mu m$.
The flux difference between IRAS 25 $\mu m$ and Spitzer MIPS 24 $\mu m$ is negligible \citep{2009ApJ...703.1672K} and that between Spitzer MIPS 24 $\mu m$ and WISE W4 is around 16 per cent \citep{2014MNRAS.443.1329H}.
Thus, we adopt a relation $L_{25\,\mu m}=1.19 \times L_{22\,\mu m}$ to convert WISE W4 luminosity to IRAS 25 $\mu m$ luminosity.

The comparison between GSWLC-M2 SED SFRs and NUV plus MIR SFRs derived above is presented in the left panel of Fig. \ref{fig:s1} where both SFRs are divided by stellar masses in GSWLC-M2.
Above SED sSFR $\sim10^{-11}\,\mathrm{yr}^{-1}$, the relation is very close to a 1:1 relation (black dashed line) but it deviates strongly below this threshold with NUV plus MIR sSFRs saturated and SED sSFRs extending further toward lower values.
This has made the linear fitting with orthogonal distance regression method give a super-linear relation (green dashed line), while the general dispersion is not large (0.25 dex).
Considering that toward low sSFR, an increasing fraction of SED SFRs are derived without detected FUV flux in SED fitting, here we test if the inclusion of detected FUV (and also explicitly for NUV and W4), for which case the results of SED fitting should be more accurate, can ease the tension between SED SFRs and NUV+MIR SFRs.
The result is shown in the right panel of Fig. \ref{fig:s1}.
Due to the requirement of FUV detection, galaxies on this plane mainly populate the high sSFR part.
Indeed the incorporation of FUV information reduces the scatter.
However, below sSFR $\sim10^{-11}\,\mathrm{yr}^{-1}$ galaxies deviate from a 1:1 relation in the same manner as in the left panel.
This suggests that the discrepancy is not due to data quality but intrinsic, and a reasonable explanation is that the coefficients in NUV+MIR SFRs are fixed while it is possible that the contribution to NUV and MIR fluxes from young stars change significantly for galaxies with different star formation levels \citep{2016A&A...591A...6B}.

\begin{figure}
	\begin{center}
		\includegraphics[width=0.45\textwidth]{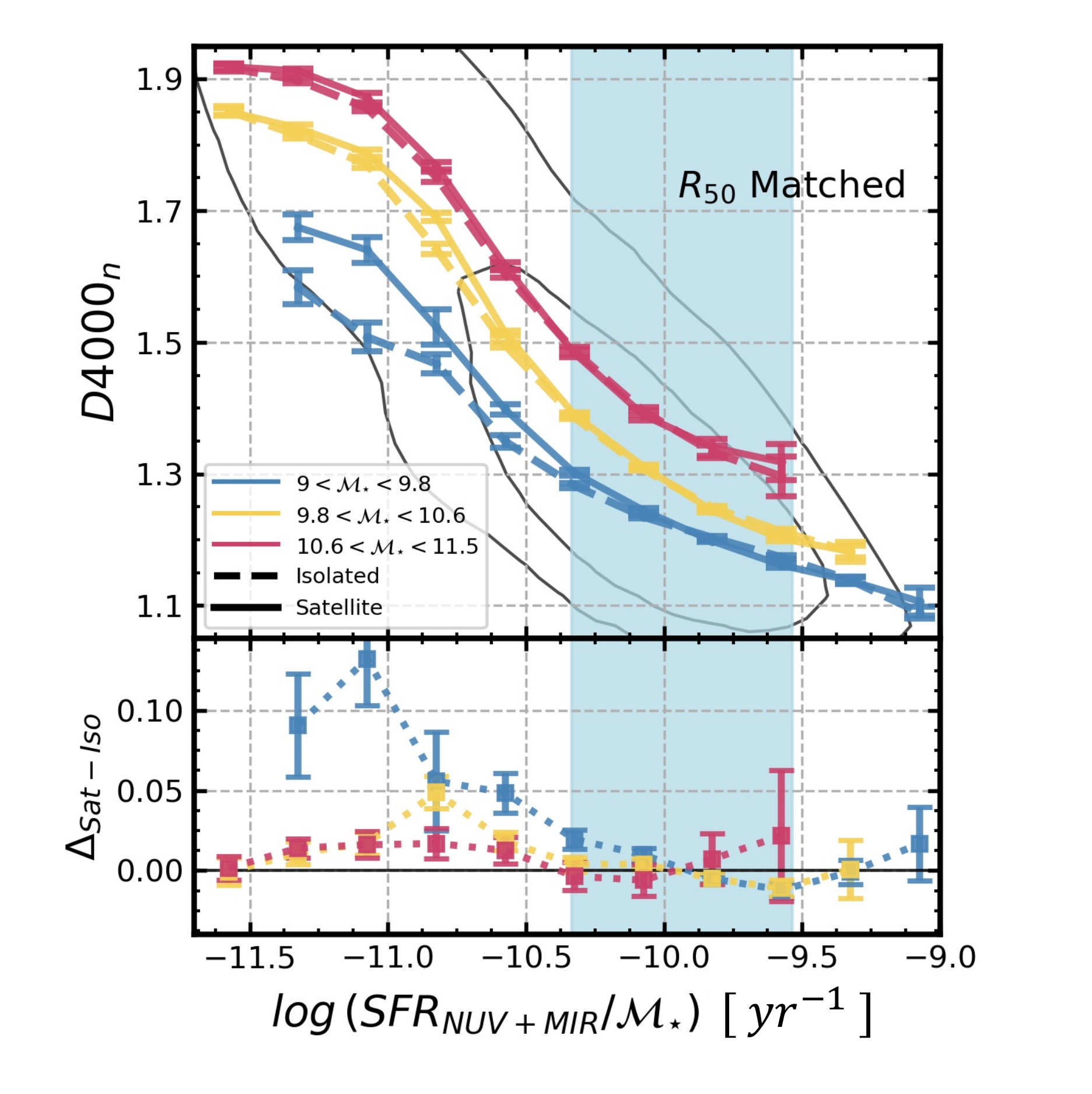}
		\caption{Reproduction of the left part of Fig. \ref{fig:100bs1} using SFRs derived from GALEX NUV plus WISE W4 luminosities.
			}
		\label{fig:s2}
	\end{center}
\end{figure}

Despite the large systematics at low sSFR regime, here in Fig. \ref{fig:s2} we reproduce the main signal shown in the left panel of Fig. \ref{fig:100bs1} using NUV+MIR sSFRs.
The central \dfourk\ excess at a given total sSFR of galaxies in the lowest mass bin still increases below the SFMS and reaches a level of around 0.1.
This shows that our main conclusion does not depend on the choice of SFR recipe.

%%%%%%%%%%%%%%%%%%%%%%%%%%%%%%%%%%%%%%%%%%%%%%%%%%%%%%%%%%%%%%%%%%%%

%%%%%%%%%%%%%%%%%%%%%%%%%%%%%%%%%%%%%%%%%%%%%%%%%%%%%%%%%%%%%%%%%%%%%%%%%%%%%%

% Don't change these lines
\bsp	% typesetting comment
\label{lastpage}
\end{document}